\documentclass[aps,prb,twocolumn,showpacs,amsmath,amssymb,showpacs]{revtex4-1}
 \usepackage{graphicx,picture,calc}
\usepackage{lipsum}
\usepackage{hyperref}
\usepackage{subfigure}
\usepackage{braket}
\usepackage{bm}
\usepackage{color}
\usepackage{float}
\usepackage[babel=true]{csquotes}

\graphicspath{ {FiguresDWs/} }

\newcommand{\rem}[1]{}

\newcommand{\fsav}[1]{\left< #1\right>_{\hat p}}
\newcommand{\lw}{\linewidth}
\newcommand{\beq}{\begin{equation}}
\newcommand{\eeq}{\end{equation}}
\newcommand{\beqa}{\begin{eqnarray}}
\newcommand{\eeqa}{\end{eqnarray}}

\newcommand{\refe}[1]{\eqref{#1}}
\newcommand{\refE}[1]{Eq.~\eqref{#1}}

\newcommand{\tr}{\mathrm{Tr}}

\newcommand{\im}{\mathrm{Im}}
\newcommand{\grad}{\bm \nabla}

\newcommand {\Ndw}{ {N_{\rm DW}}}
\newcommand{\abs}{{\rm \tiny ABS}}

\newcommand{\fflo}{{\rm FFLO}}
\newcommand{\dw}{{\rm DW}}
\newcommand{\No}{{\rm N}}

\newcommand{\imp}{{\rm imp}}

\newcommand{\hp}{{\hat p}}
\newcommand{\vR}{\bm R}

\newcommand{\vj}{\bm j}

\def\pder#1#2{\mbox{$\displaystyle\frac{\partial #1}{\partial #2}$}} 

\begin{document}
\title{Heat transport in nonuniform superconductors}

\author{Caroline Richard} 
\author{Anton B. Vorontsov}
\affiliation{Department of Physics, Montana State University, Bozeman, Montana 59717, USA}

\begin{abstract}
We calculate electronic energy transport in inhomogeneous
superconductors using a fully  self-consistent non-equilibrium quasiclassical
Keldysh approach. We develop a general theory and apply it a superconductor with 
an order parameter that forms domain walls, of the type encountered in 
Fulde-Ferrell-Larkin-Ovchinnikov state. 
The heat transport in the presence of a domain wall is inherently anisotropic and non-local. 
Bound states in the nonuniform region play a crucial role and control 
heat transport in several ways: (i) they modify the spectrum of quasiparticle states and result in 
Andreev reflection processes, and (ii) they hybridize with impurity band and produce 
local transport environment with properties very different from those in uniform superconductor. 
As a result of this interplay, heat transport becomes highly sensitive to temperature, magnetic field and 
disorder. For strongly scattering impurities we find that the transport \emph{across} domain walls at low temperatures 
is considerably more efficient than in the uniform superconducting state. 
\end{abstract}

\date{\today}

\maketitle

 
\section{introduction}
 
Electronic heat transport is a powerful tool to explore properties of the superconducting state. 
It is a bulk probe, 
that encodes information about both  density of electronic states and  
quasiparticle relaxation times. 
Heat conductivity experiments have been used extensively to study structure and symmetries 
of the superconducting order parameter in many different compounds.\cite{MatsudaVekhterReview}
The low-temperature behavior of thermal conductivity 
is a signature of either the absence or presence of low-energy excitations.\cite{Graf1996} 
It can also be used as a directional probe of the gap structure, since it depends on the 
velocity of the low-energy excitations. 
One can measure the anisotropy of thermal conductivity along different directions and 
identify the Fermi velocity vectors of nodal quasiparticles.\cite{LussierPRB1996,NormanHirschfeld1996,Vekhter2007thcond2D}
Another way to study the nodal structure 
is to observe the response of nodal excitations to a rotated magnetic field.\cite{MatsudaVekhterReview} 
The external magnetic field modifies the density of states\cite{Volovik1993,Vekhter2001thermodyn} 
and the quasiparticles scattering times.\cite{Franz1999,VekhterHoughton1999,Vorontsov2007thtransp} 
The magnitude of these effects depends on the orientation 
of the magnetic field relative to the nodes of the order parameter and the 
direction of the heat flow.\cite{MatsudaVekhterReview}  

The power of this technique, however, is also the reason why the 
interpretation of thermal transport measurements 
is a  difficult task, since  density of states and transport time of 
quasiparticles may not be independently available. 
In this respect, experiment and theory must be employed together in the analysis of 
data for reaching definite conclusions. 
In uniform superconductors heat conductivity has been investigated in great details, 
using several approaches: Boltzmann transport theory,\cite{MineevSamokhin}
Green's functions technique,\cite{Hirschfeld1988}
and quasiclassical methods,\cite{Graf1996}
that prompted rapid development on the experimental side. 
 
There is growing interest in using thermal transport to study 
nonuniform superconductors\cite{Capan2004fflo} and topological surface states.\cite{Ryu2012,Nakai2014,sothmann2016fingerprint} 
However, from the theory side,  little is known about heat flow in the presence of a spatially-varying
order parameter.  The challenge here is to understand how quasiparticles
transport energy from one point to another when both the quasiparticle density of 
states and scattering mean free path  depend on both energy and
position. Under these conditions it is important to treat on the same footing 
Andreev particle-hole conversion processes in
inhomogeneous regions\cite{Andreev:1964wk,*Andreev:1965vc} 
and  scattering processes on impurities.   

As a result, in nonuniform superconductors  calculation  of heat transport 
is difficult and  so far  has been carried out 
only in two different approximations. 
In the strongly inhomogeneous situation, 
as in the case of periodic and moderately dense Abrikosov vortex lattice near $H_{c2}$, one can 
average over vortex lattice unit cell,\cite{Maki1967,BPT1967} assuming local formula 
relating heat current to the temperature gradient, 
$\vj_h(\vR) = -\hat\kappa \; \grad T(\vR)$, to hold everywhere. 
In this approach one can analyze the effects of disorder and magnetic field on 
density of states, lifetime 
and heat transport of spatially extended quasiparticles 
outside vortex cores.\cite{VekhterHoughton1999,Vorontsov2007thtransp} 
In a very different setting, the heat transport through a pinhole supporting
Andreev bound states (ABS)  was investigated \cite{ZhaoPRL2003,*ZhaoPRB2004}. 
When a phase bias $\varphi$ applied across the pinhole, 
highly degenerate Andreev bound states\cite{Andreev:1964wk,Andreev:1965vc} appear 
at subgap energies controlled by both $\varphi$ and the transparency of the pinhole. 
The sudden  temperature drop across the pinhole produces local heat current that depends on the 
phase bias, $ j_h = -\kappa(\varphi) \delta T$. The bound states lying at subgap
energy do  not directly couple to the continuum of quasiparticles to transport
heat. Nevertheless, their presence modifies the effective transparency of the
pinhole for quasiparticles above the gap.  In particular, for a pinhole with
perfect transparency, the subgap bound states reduce locally the spectral weight of
continuum states which suppresses the heat flow, $\kappa(\varphi ) < \kappa(0)$.
By contrast, at low transparency, the ABS lie just below the gap edge and  enhance heat conductivity, $\kappa(\varphi ) >
\kappa(0)$, due to   a resonance  with the continuum.\cite{ZhaoPRL2003}  However, in topological insulator junctions, the zero energy ABS are topologically protected, preventing  such resonance. \cite{sothmann2016fingerprint} 

Both of these approaches have limited applicability.  
In the pinhole calculation the sudden drop approximation means point-localized
ABS and lack of impurity scattering effects.
The averaging procedure, on the other hand, works well 
for high temperature and fields where vortices are dense, 
but less well at low temperatures and fields, 
and even worse  with  fully  gapped superconductors. 
It relies on a presence of significant number of 
spatially extended low-energy quasiparticles, but has no way 
of including the contribution of localized vortex core states. 
An exact theoretical treatment of the 
thermal transport that simultaneously takes into account impurity scattering in 
spatially varying order parameter landscape, effects 
of spatially localized Andreev bound states, and position-dependent density of states, 
is lacking. 
In this direction,  we provide,  in this paper, a basis for future explorations of general nonuniform 
superconducting configurations. 
 
There are several important details that we include in 
complete treatment of the problem. 
First, effects of the Andreev states localized in the inhomogeneous region is 
taken into account on the 
same footing with the effects of the impurities that also produce midgap Andreev states distributed throughout the sample. 
Both kind of bound states result in modifications in the quasiparticle spectrum and scattering time, 
and their interaction is important. 
Second, the broken translation and rotation symmetry that appear in systems with a spatially modulated order parameter 
in general result in additional `vertex corrections' to the transport lifetime.\cite{Eschrig1999vortices,Eschrig2002vortices} 
Third, the mean free paths of the quasiparticles can be longer than the
coherence-length scale of the order parameter variations, 
thus invalidating the picture of a local equilibrium and local response 
even for small temperature gradients. 

As a particular model for the inhomogeneous phase we consider a domain wall between 
two degenerate configuration of the order parameter that changes sign across the 
wall, $\Delta(-\infty) = -\Delta(+\infty)$. We also consider a more complicated 
configurations with periodic collection of multiple domain walls. 
We enforce the order parameter modulation through boundary conditions on the edges of the sample, 
and self-consistently compute the  profile of the domain walls and spatially-dependent impurity self-energies.
The domain walls have width of several coherence lengths, 
and   host   highly degenerate zero-energy Andreev bound states. 
Such profile of the order parameter is a realization of 
Larkin-Ovchinnikov configuration of the speculative Fulde-Ferrell-Larkin-Ovchinnikov (FFLO) phase. 
\cite{Larkin:1964uu,*Larkin:1965wj,Burkhardt:1994ts,VorontsovFFLO}
Building a theoretical understanding of the thermal conductivity in this 
phase is important for experimental attempts to detect spatial modulations of the 
order parameter using heat transport.\cite{Capan2004fflo} 
At this point it is not known how anisotropic is the heat flow in the presence of the 
hypothetical FFLO domain wall structures in the order parameter. 
For example, the typical assumption is that the heat flow across the 
domains is strongly constricted due to the presence of the ABS 
that do not carry heat. 
We find that this is not, in fact, true, and the combination of 
impurity scattering effects with spatially-extended range of the bound states
can produce both suppression and enhancement of thermal transport 
compared to the uniform state.



The organization of the paper is as follows. 
In section \ref{sec:model} we describe our model and relate it to the experimental 
measurement technique. In section \ref{sec:formalism} we 
develop  the formalism to  compute  thermal transport in nonuniform superconductor 
using Keldysh quasiclassical approach and t-matrix treatment of disorder. 
The linear response is discussed in section \ref{sec:lr}; 
in \ref{sec:bc} we formulate the boundary conditions 
for the quasiclassical propagator amplitudes. 
We apply this technique to compute heat flow across a single 
domain wall in section \ref{sec:singleDW}, 
and in \ref{sec:DWs} we generalize it to the case of multiple domain walls 
and investigate heat transport dependence on number of domains, temperature, disorder
and spacing of domain walls. 
Finally, in section \ref{sec:zeeman}, we address the
modifications arising  from the Zeeman shifts of quasiparticle energies.

\section{Model\label{sec:model}}
 
\begin{figure}[t!]
\includegraphics[width = 0.9\linewidth]{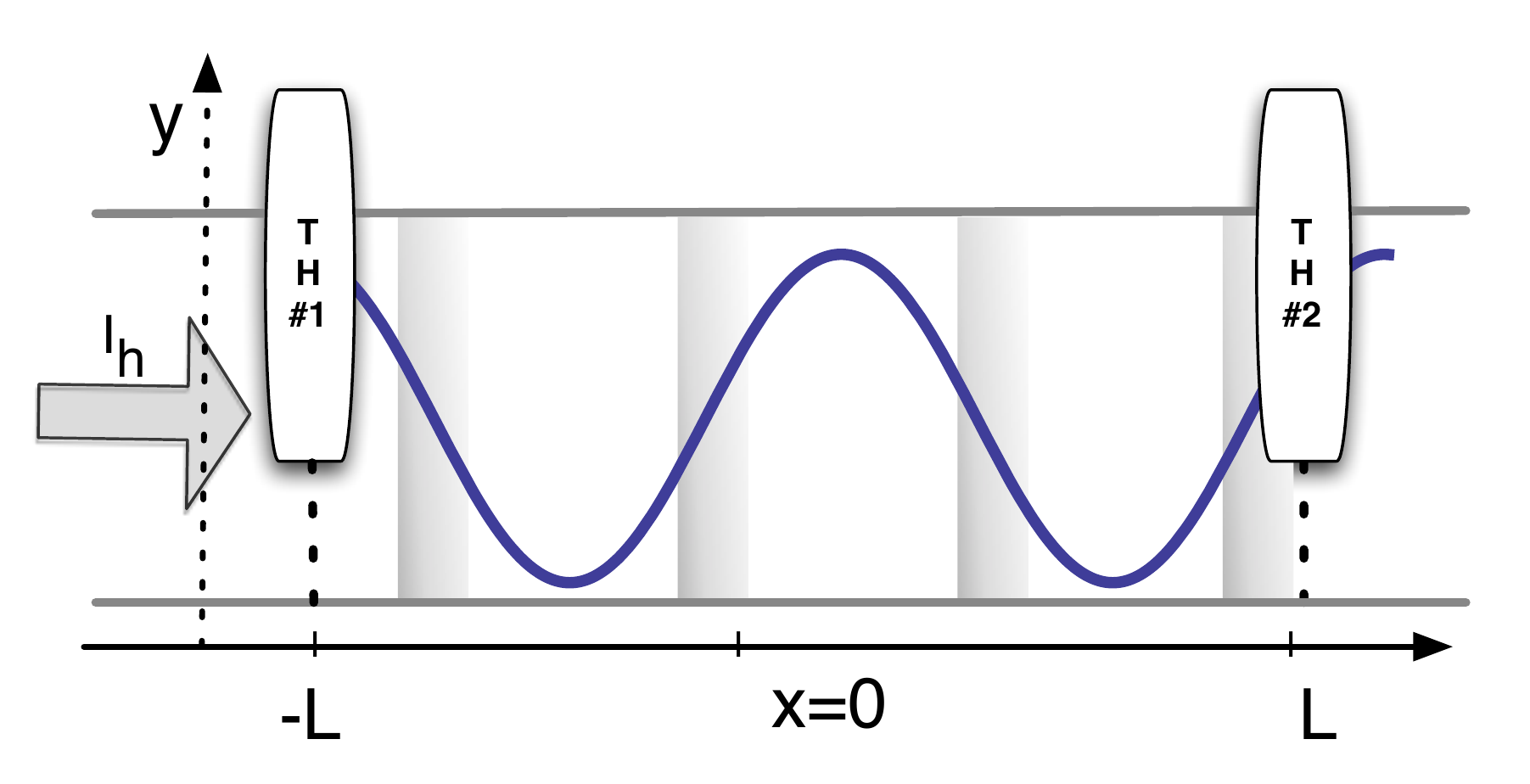}
\caption{ \label{fig:model} 
A typical experimental setup for heat conductivity measurement 
involves establishing a steady state current in the sample and 
measuring 
the effective local temperature $T_1$ and $T_2$.  Heat conductivity is defined  $\kappa= I_h/(T_1-T_2)$. 
We set a two-dimensional superconductor 
in the $xy$ plane and for a given heat current $I_h$ along the $x$-axis 
we calculate temperature difference between points TH \#1 and \#2 at $x=\pm L$. 
The sin line is a schematic representation of an FFLO
modulation of the  order parameter with shaded regions representing domain walls. 
}
\end{figure}

We consider a spin-singlet superconductor with quasi two-dimensional cylindrical Fermi surface.
We will focus on the case of a $d$-wave superconductor. 
The d-wave order parameter has nodes in momentum space,
$\Delta(\hp) \propto \cos 2(\phi_\hp-\alpha)$.   
Impurities are randomly distributed throughout the sample with concentration $c_{\rm imp}$. 
The impurity scattering potential is assumed point-like and isotropic with
amplitude $u$. 
System is set out of equilibrium by introducing a thermal current flowing 
along the $x$-axis. As the stationary state is reached a temperature gradient
builds up. We define the heat
conductivity between two points as $ \kappa= I_h \times 2 L/(T_1-T_2)$ where $I_h$ is the
stationary-state heat flow, and $T_1$ and $T_2$ are the local temperatures as if
measured by two distant thermometers positioned at $x=\pm L$. Our goal is to
compute the effective temperature bias  $dT= T_1-T_2$ for a given $I_h$, 
in the presence of spatially modulated order parameter as shown in 
Fig.~\ref{fig:model}.      

\section{Theory\label{sec:formalism}}

A convenient approach to study nonuniform superconductors out of 
equilibrium is the quasiclassical formulation of the Keldysh 
technique.\cite{Serene:1983vc}
It is formulated in terms of the Green function 
$\hat g(\vR, \hp, \epsilon)$, which for  stationary states depends on the 
center of mass coordinate $\vR$, direction of the relative momentum on
the Fermi surface $\hp$, and the energy $\epsilon$.  
It is a $8\times8$ matrix in particle-hole (Nambu), spin, and Keldysh 
space. 
In the Keldysh space, it is given  by 
\beq 
\hat g= \left(\begin{array}{cc}
g^R & g^K \\ 0 & g^A 
\end{array}\right) \,, 
\label{eq:kel}
\eeq 
where the superscripts $R/A/K$ stand for Retarded, Advanced and Keldysh. 
The Retarded and Advanced components $g^{R/A}$ carry information about 
density of states and correlations, while $g^K$ encodes the quasiparticles' dynamics 
and distribution function. 
Each of the three components are $4\times 4$ matrices, parametrized by outer products of 
$2\times 2$ Pauli matrices in spin and particle-hole spaces $\sigma_i \otimes \tau_j$ ($i,j=x,y,z$). 

We use the quasi-classical propagator to compute 
the density of electronic states (DoS) $N(\vR,\hp, \epsilon)$ 
\beq
\frac{N(\vR,\hp, \epsilon)}{N_F}= 
- \frac{1}{\pi} \im \left[ \frac14 \tr\left[ \tau_zg^R(\vR, \hp, \epsilon) \right] \right] \,,
\label{eq:DOSdef}
\eeq
and local heat current $\vj_h(R)$, and its spectral density $\vj_h(\vR,\epsilon)$, 
\begin{align}
\begin{split}
\vj_h( \vR)=  2 N_Fv_F\int\limits_{-\infty}^{+\infty} \frac{d\epsilon}{4\pi i} \int d\hp\; 
[\epsilon \, \hp] \; 
\frac14 \tr\left[g^K(\vR,\hp,\epsilon) \right] \,
\\
\equiv \int\limits_{-\infty}^{+\infty}d\epsilon\,
\vj_h(\vR,\epsilon) \,.
\end{split}
\label{eq:heatqdef}
\end{align}
Here $N_F$ is density of state at Fermi energy per spin in the normal metallic state, 
$v_F$ is the Fermi velocity, and $\tr=\tr_4$ is the trace operator over spin and Nambu space.  
$\int d\hp\; \dots = \fsav{\dots} = \int \frac{d\phi_\hp}{2\pi} \dots$ is the normalized 
Fermi surface integral. 
We note that to write the heat current as 4-trace over spin \emph{and} particle-hole space 
instead of usual spin-trace over just upper left component of $g^K$ we used symmetry of 
the Keldysh component
$ g^K(\vR,\hp,\epsilon)^{tr} = \tau_y g^K(\vR,-\hp,-\epsilon) \tau_y$.\cite{Serene:1983vc}

The quasiclassical Green function $\hat g$ is normalized 
\beq
\hat  g^2(\vR,\hp, \epsilon) =-\pi^2 \,,\qquad 
\tr [ g^{R,A} ]=0 \,,
\label{eq:EilNorm}
\eeq
and satisfies the Eilenberger equation 
\beq
[\epsilon \hat\tau_z - \hat\sigma, \hat g]+ i v_F\hp\cdot\grad \hat g=0 \,,
\label{eq:Eil}
\eeq
where $8\times8$ self-energy $\hat \sigma(\vR, \hp, \epsilon)$ has the same structure in 
Keldysh space as \refE{eq:kel} and $\hat\tau_z = \mbox{diag}(\tau_z, \tau_z)$. 
The retarded and advanced components for 
singlet superconductor are parametrized as follows (${\rm x}=R,A$):
\beq
\sigma^{\rm x} =
\left(\begin{array}{cc}\Sigma^{\rm x} & \Delta^{\rm x}(i\sigma_y) 
\\ 
(i\sigma_y) \tilde\Delta^{\rm x} & \tilde \Sigma^{\rm x}\end{array}\right) \,,
\eeq 
and the Keldysh part is 
\beq
\sigma^K =
\left(\begin{array}{cc}\Sigma^K & \Delta^K(i\sigma_y) 
\\ 
- (i\sigma_y) \tilde\Delta^K & - \tilde \Sigma^K \end{array}\right) \,.
\eeq 
Components of these matrices are related to each other 
through symmetries\cite{Serene:1983vc} defined by the $\tilde{\hspace{4pt}}$-operation 
that reverses momentum and energy of complex-conjugated quantities, e.g.
$\tilde \Delta^{\rm x}(\vR,\hp,\epsilon) = \Delta^{\rm x}(\vR,-\hp,-\epsilon)^*$.
Diagonal self-energy terms $\Sigma,\tilde\Sigma$ are due to  impurity scattering effects. 
The off-diagonal terms contain the mean field order parameter and impurity contributions:
$$
\Delta^{R/A}(\vR, \hp, \epsilon) = \Delta(\vR, \hp) + \Delta_{\imp}^{R/A}(\vR, \epsilon)
$$
while the Keldysh mean fields identically zeros, leaving only impurity contributions
$$
\Delta^{K}(\vR, \epsilon) = \Delta_{\imp}^{K}(\vR, \epsilon) \,.
$$
The mean-field order parameter is computed self-consistently from 
\beq
\Delta (\vR, \hp) = \int\limits_{-\epsilon_c}^{+\epsilon_c} 
\frac{d\epsilon}{4\pi i} \int d\hp' \; V(\hp, {\bm \hat p'}) \; f^K(\vR, \hp,  \epsilon) \,.
\label{eq:scDelta} 
\eeq 
We consider pair potential $V(\hp, \hp')= V_0 \; {\cal Y}(\hp) {\cal Y^*}(\hp')$ 
and $f^K = \frac14 \tr[ \frac{\tau_x+i\tau_y}{2} (-i\sigma_y) g^K]$ is the upper-right singlet component 
of the Keldysh Green's function. 
The momentum space basis functions are 
${\cal Y}(\hp)= 1$ for $s$-wave, and  
${\cal Y}(\hp)= \cos{2(\phi_\hp-\alpha)}$ for $d$-wave. 
The cut-off energy $\epsilon_c$ and the interaction amplitude $V_0$ are 
eliminated, in the usual manner, in favor of clean-case transition temperature $T_c$. 

The impurity self-energy part is self-consistently determined within
the $\hat t$-matrix approximation. For   randomly distributed isotropic
scattering centers, the $8\times8$ $\hat t$-matrix equation is 
$\hat t = u \hat 1 + u \fsav{\hat g} \hat t$ 
gives self-energy $\hat \sigma_\imp= c_\imp \hat t$. 
Using traditional definitions of impurity scattering rate 
$\Gamma= {c_\imp}/{\pi N_F }$ and phase shift 
$\tan\delta= u\pi N_F $ in terms of impurity concentration $c_{\rm imp}$ 
and amplitude $u$ of the point-like scattering potential, 
the $4\times4$ self-energy matrices are determined from (${\rm x}=R,A$)
\begin{align} 
\begin{split} 
& \sigma_\imp^{\rm x}(\vR,\epsilon) 
= \Gamma\tan \delta+\tan\delta \fsav{\frac{g^{\rm x}(\vR,\hp,\epsilon)}\pi} \sigma^{\rm x}_\imp(\vR,\epsilon) \,,
\\ 
& \sigma_\imp^K(\vR,\epsilon) = \frac1\Gamma \sigma^R_\imp(\vR,\epsilon) 
\fsav{ \frac {g^K(\vR, \hp, \epsilon)}{\pi}} \sigma^A_\imp(\vR,\epsilon) \,.
\end{split} 
 \label{eq:scImp} 
\end{align} 
In the following, we will mainly make comparison between 
the Born ($\delta \to 0$) and Unitary ($\delta=\pi/2$) limits. 
Note that in the absence of inelastic scattering, 
the self-consistent calculation of impurity 
$\sigma_{\rm imp}$ and order parameter $\Delta$ that includes non-equilibrium effects, 
guarantees 
conservation of energy and charge. 
In particular, self-consistent calculation of self-energies 
including corrections due to the heat flow will automatically satisfy 
${\rm div}\vj_h(\vR,\epsilon)=0$ 
- condition of no energy accumulation, see Appendix \ref{app:conservation}.
 
The solution of the self-consistent system of coupled equations for nonuniform
states is most conveniently obtained 
using Riccati parametrization\cite{Eschrig2000}. 
The retarded ($-$ sign) and advanced ($+$ sign) Green's functions are given by 
\beq 
g^{\rm x}=\frac {\mp i \pi }{1 + \gamma^{\rm x} \tilde \gamma^{\rm x}}\left(
\begin{array}{cc} 
1-\gamma^{\rm x} \tilde\gamma^{\rm x}  &  2\gamma^{\rm x} ( i \sigma_y )
\\ 
- 2\tilde \gamma^{\rm x} (i \sigma_y )  &  -(1-\tilde \gamma^{\rm x}\gamma^{\rm x}) 
\end{array}\right) \,, 
\label{eq:RicRA}
\eeq
where $\gamma^{{\rm x}}(\vR, \hp , \epsilon)$ are retarded/advanced scalar coherence functions that
are zeros in  bulk normal state. 
The Keldysh component takes the form
\begin{align} 
\begin{split} 
& g^{K}=\frac {-2 \pi i  }{(1 + \gamma^R \tilde \gamma^R)(1 + \gamma^A \tilde \gamma^A)} \times 
\\
& \left( \begin{array}{cc} 
x^K+ \gamma^R \tilde x^K \tilde \gamma^A & - (\gamma^R \tilde x^K -x^K\gamma^A) i\sigma_y  \\
-i\sigma_y (\tilde \gamma^R x^K-\tilde x^K \tilde \gamma^A)& \tilde x^K + \tilde \gamma^R x^K  \gamma^A
\end{array}\right) \,,
\end{split} 
\label{eq:RicK}
\end{align} 
where $x^K(\vR, \hp, \epsilon)$ is the (scalar) distribution function. 
We explicitly took out the singlet spin dependence 
of coherence amplitudes, compared with Ref.~\onlinecite{Eschrig2000}, 
where the coherence functions are spin matrices. 
Coherence and distribution amplitudes are related to each other 
through $\tilde{\hspace{4pt}}$-relation, as well as\cite{Eschrig2000}
\begin{align} 
\begin{split} 
& \gamma^A(\vR, \hp, \epsilon) = -\tilde \gamma^R(\vR, \hp, \epsilon)^* \,,
\\ 
& x^K(\vR, \hp, \epsilon) = x^K(\vR, \hp, \epsilon)^* \,.
\end{split} 
 \label{eq:symmRic} 
\end{align} 
The distribution function is not unique, and the usual choice in equilibrium is 
\beq
x^K_{0} = \Phi_0(\epsilon/T) (1+\gamma^R\tilde\gamma^A) \,, 
\label{eq:xKeq}
\eeq
where
$$
\Phi_0\left({\epsilon}/{T}\right) = \tanh({\epsilon}/{2T}) = 1-2f(\epsilon/T)
$$ 
and 
$f(\epsilon/T)=\left[\exp(\epsilon/T)+1\right]^{-1}$ 
is the Fermi distribution at temperature $T$.
Transport-like equations for coherence and distribution functions follow from \refE{eq:Eil}
\begin{align} 
\begin{split} 
& 
i v_F \hp \cdot \grad \gamma^{\rm x} + (2\epsilon-\Sigma^{\rm x}+\tilde\Sigma^{\rm x}) \gamma^{\rm x} 
+ \tilde\Delta^{\rm x} \, (\gamma^{\rm x})^2 + \Delta^{\rm x} = 0 \,, 
\\ 
& 
i v_F \hp \cdot \grad x^K + (\gamma^R \tilde\Delta^R -\Sigma^R + \Delta^A \tilde\gamma^A + \Sigma^A) x^K  
= 
\\ 
&\hspace{3cm} \gamma^R \tilde\Sigma^K \tilde\gamma^A - \Delta^K \tilde\gamma^A - \gamma^R \tilde\Delta^K - \Sigma^K
\,.
\end{split} 
 \label{eq:Rictr} 
\end{align} 

\subsection{Linear response}
\label{sec:lr}

In the absence of heat current, $ \vj_h=0 $, the system is assumed in global equilibrium
at temperature $T$, 
with $x^K= x^K_0(\vR,\hp,\epsilon)$ given by \refE{eq:xKeq},
and equilibrium coherence functions $\gamma^{\rm x}=\gamma_0^{\rm x}(\vR,\hp,\epsilon)$ 
found from 
\beq
i v_F \hp \cdot \grad \gamma_0^{\rm x}+ (2\epsilon-\Sigma_0^{\rm x}+\tilde\Sigma_0^{\rm x}) \gamma_0^{\rm x} 
+ \tilde\Delta_0^{\rm x} \, (\gamma_0^{\rm x})^2 + \Delta_0^{\rm x} = 0 \,. 
\label{eq:gamma}
\eeq
In uniform superconductors, $\grad \gamma_0^{\rm x}(\vR, \hp, \epsilon)=0$
and the solution of Eq.~(\ref{eq:gamma}) for the retarded coherence function is 
\beq
\gamma_{u}^R(\hp, \epsilon)=- \frac{\Delta_u^R}
{{\bar\epsilon}^R+i \sqrt{ \Delta_u^R\tilde\Delta_u^R-(\bar{\epsilon}^{R})^2 }},\label{eq:gammaU}
\eeq
with $\bar \epsilon= \epsilon - (\Sigma^R_u-\tilde \Sigma^R_u)/2$. 
In the following, the subscript \enquote{$u$} stands for `uniform', 
and subscript \enquote{$0$} will refer to the equilibrium solution. 

In the presence of a small heat current $\vj_h \neq 0$ that is assumed to be 
time-independent (stationary state), 
the system is out-of-equilibrium. 
In  linear response, 
we expand coherence and distribution functions around their equilibrium values 
\begin{align}
\begin{split}
\gamma^{{\rm x}}(\vR, \hp, \epsilon) =& \gamma_0^{\rm x}(\vR, \hp, \epsilon)+\gamma^{\rm x}_1(\vR, \hp, \epsilon) \,, 
\\
x^K(\vR, \hp, \epsilon)  =& \Phi_0(\epsilon)(1+ \gamma^R_0 \tilde\gamma^A_0) 
\\
&+ \Phi_0(\epsilon) \; (\gamma_0^R\tilde\gamma_1^A+\gamma_0^R\tilde\gamma_1^A)+ x^a \,.
\end{split}
\label{eq:xK_exp}
\end{align}
The deviation of the distribution function from equilibrium, $x^K-x^K_0$,
is described by two terms. 
The first accounts for change in the density of states through corrections to the retarded and advanced 
functions, it is weighted by the equilibrium Fermi distribution $\Phi_0(\epsilon)$. 
The second term, $x^a(\vR, \hp, \epsilon)$, is the anomalous, or dynamical distribution function. 
It determines the dynamical part of the Keldysh Green's function 
$ \hat g^a = \hat g^K -  \hat g^K_0 - \Phi_0 (\hat g_1^R - \hat g_1^A) $, 
\begin{widetext}
\beq
\hat g^{a}=\frac {-2 \pi i  }{(1 + \gamma_0^R \tilde \gamma_0^R)(1 + \gamma_0^A \tilde \gamma_0^A)}
\left( \begin{array}{cc} 
x^a+ \tilde x^a \gamma_0^R \tilde \gamma_0^A & (x^a\gamma_0^A-\tilde x^a \gamma_0^R) i\sigma_y  \\
-i\sigma_y (x^a\tilde \gamma_0^R-\tilde x^a \tilde \gamma_0^A)& \tilde x^a + x^a \tilde \gamma_0^R \gamma_0^A
\end{array}\right) \,. 
\label{eq:RicKga}
\eeq
In linear response the heat current depends only on the \emph{equilibrium} spectral properties through 
coherence amplitudes $\gamma^{R,A}_0$, 
and the dynamical part of distribution function, $x^a$: 
\beq 
\vj_h  = - 2 N_F v_F  \int\limits_{-\infty}^{+\infty} d\epsilon \fsav{  
\epsilon \hp \, 
\frac{ x^a(1+ \tilde \gamma_0^R\gamma_0^A) + \tilde x^a (1+\gamma_0^R\tilde \gamma_0^A)}
{4(1+\gamma_0^R \tilde \gamma_0^R)(1+\gamma_0^A \tilde \gamma_0^A)} 
}  \,.
\label{eq:qLinResp}
\eeq 
To obtain equation for $x^a$ we linearize \refE{eq:Rictr}. 
We linearize with respect to the global equilibrium at temperature $T$, where 
$\Phi_0(\epsilon/2T)$ in \refE{eq:xK_exp} is position-independent - in this case 
the linearized equation reads\cite{Eschrig2000} 
\beq
i v_F\hp \cdot\grad x^a + \frac{ iv_F} {\ell(\vR, \hp, \epsilon)} x^a 
=\gamma_0^R\tilde \gamma_0^A \tilde \Sigma^a-(\Delta^a \tilde \gamma_0^A + \tilde \Delta^a \gamma_0^R + \Sigma^a)
\label{eq:xAlin} \,, 
\eeq
with equation for the $\tilde{x}^a$-function obtained from this one by employing the definition of $\tilde{\hspace{4pt}}$-operation. 
In the above equations we introduced parameter 
\beq
\ell( \vR,\hp, \epsilon) =i v_F/(\gamma_0^R \tilde \Delta_0^R - \Sigma_0^R + \tilde \gamma_0^A \Delta_0^A+\Sigma_0^A) \,,
\label{eq:elldef}
\eeq  
that is purely real, as follows from the symmetries of the coherence functions and self-energies, 
and has dimension of length. 
In the normal metallic phase 
$\ell= v_F / 2 \Gamma \sin^2\delta = v_F \tau_N= \ell_ N $ 
matches with  the elastic mean free path. 
The dynamical self-energy entering \refE{eq:xAlin} 
as the source term 
is self-consistently computed from $x^a$ and $\gamma_0^{\rm x}$:  
\beq
\sigma^a(\vR, \epsilon) \equiv \sigma_{\rm imp}^a  \equiv \left(\begin{array}{cc} \Sigma^a&
(i\sigma_y) \Delta^a\\ - (i\sigma_y) \tilde \Delta^a&-\tilde \Sigma^a
\end{array}\right)
= \frac 1\Gamma \sigma^R_{0, \rm imp}(\vR,\epsilon) 
\fsav{ \frac{g^a(\vR, \hp, \epsilon)}{\pi}} \sigma^A_{0, \rm imp}(\vR,\epsilon) \,.
\label{eq:sigmaAsc}
\eeq
\end{widetext}
Note that the linearization scheme in \refE{eq:xK_exp} is very convenient because the
calculation of $\vj_h$ only requires the knowledge of the anomalous $x^a$, 
which itself does not depend on spectrum corrections, $\gamma^{\rm x}_1$.  
 
\subsection{Boundary conditions}
\label{sec:bc}

To solve the transport equation \refE{eq:xAlin} for the distribution function, 
one needs to provide suitable boundary conditions for initial values of $x^a$ 
at the beginning of a trajectory $v_F \hp$, and for final value of $\tilde x^a$ 
at the end of this trajectory. 

For weak links, as in reference \onlinecite{ZhaoPRL2003},
one can assume that the system of interest is connected to
large reservoirs in equilibrium at temperatures $T_{1,2}$. 
Then, for a given quasiparticle trajectory, 
one can take equilibrium values of the coherence and dynamical amplitudes 
in those reservoirs as initial values.  

Such assumption seems inadequate to compute heat transport in bulk samples. 
Instead, given a stationary conserved heat flow in the entire sample,  
we will construct the Riccati amplitudes at $x = \pm L$ in a way that is consistent with \refE{eq:xAlin}, 
and would give a fixed thermal current 
\beq
\vj_h(\pm L)= \vj_h=\vj^{\rm BC}_h
\label{bc:current}
\eeq   
in a uniform superconducting state, away from the inhomogeneous region. 

\begin{figure}[t!]
\includegraphics[width = 0.8\linewidth]{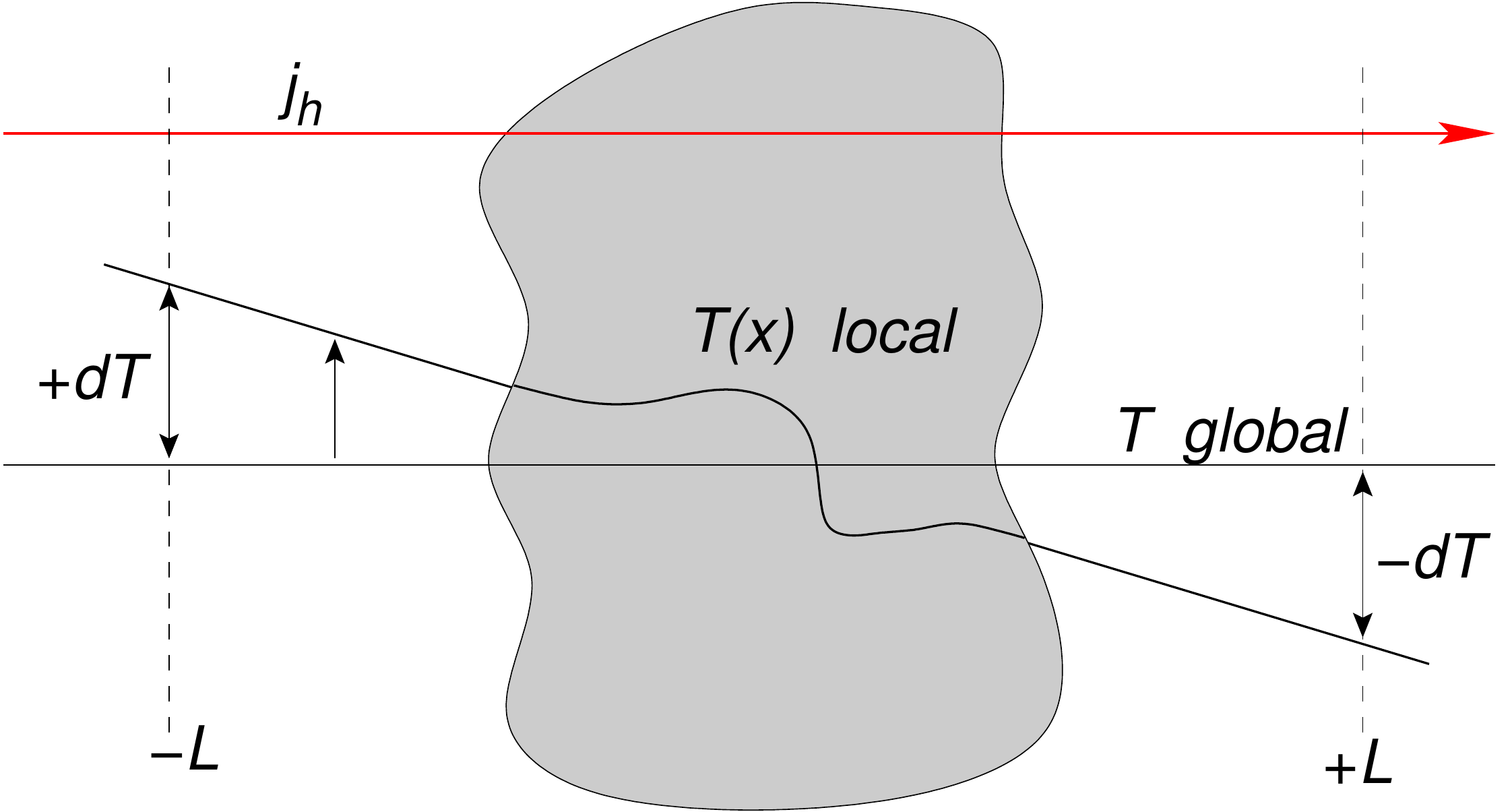}  
\caption{\label{fig:loc_glo} 
Local versus global equilibrium picture. 
The system is driven out of equilibrium by a steady uniform heat current $j_h$.  
The local equilibrium picture assumes that,
when heat flows,  a local temperature, $T(x)$,  can be defined, and its gradient 
determines the magnitude of $j_h$.
It is typically used in uniform-state problems, and we use it here to define 
boundary conditions for distribution functions away from the nonuniform (shaded) region 
of the order parameter.
Our main approach, however, is to 
expand the propagators around some global equilibrium value of the temperature $T$:  
$g(x)= g_{\rm eq}(T)+ g_c(x)$ where $g_c(x)$ determines current $j_h$.  
}
\end{figure}

To write such boundary condition we start by making a trivial observation that 
the linearization procedure that we followed, can be used to find \emph{equilibrium} functions 
at slightly different temperature $T+\delta T$. For example, the distribution function can be 
written as 
\beq
x^K_{eq}(T+\delta T) = [1+\gamma^R_0(T+\delta T) \tilde \gamma^A_0(T+\delta T)] 
\Phi_0\left(\frac{\epsilon}{2(T+\delta T)}\right) 
\nonumber
\eeq
where we show only the temperature argument explicitly. 
Decomposition \refE{eq:xK_exp} in this case gives the anomalous contribution as 
\beq
x^a_{eq}(T+\delta T) = [1+\gamma^R_0(T) \tilde \gamma^A_0(T)] \pder{\Phi_0}{T} \delta T \,,
\label{eq:xAeq}
\eeq
with $\partial \Phi_0/\partial T = - \epsilon / [ 2T^2 \cosh^2(\epsilon/2T)]$.
This $x^a_{eq}$ distribution function, on the other hand, also satisfies \refE{eq:xAlin} with appropriately determined 
self-energy through \refE{eq:sigmaAsc} that can be brought to the form 
$\sigma^a(T) = (\sigma^R_{0,imp}-\sigma^A_{0,imp}) (\partial \Phi_0 /\partial T) \delta T$.
Far from the region of spatially varying order parameter, the equilibrium functions, $\gamma^{R/A}_0(\hp,\epsilon)$ and 
$x^K_0(\hp,\epsilon)$ take their uniform values $\gamma^{\rm x}_0=\gamma_u^{\rm x}$, 
that determine the scattering length $\ell_u$ through \refE{eq:elldef}, 
and in equilibrium $\grad x^a_{eq} \propto \grad T = 0$. 

When a stationary  thermal current flows, a local temperature gradient builds up and $\delta T(\vR) = T(\vR) - T$ 
is a function of
position, see Fig. \ref{fig:loc_glo}. As a result, \refE{eq:xAeq} with local $\delta T(\vR)$ is no longer a solution to
\refE{eq:xAlin}. However, one can modify $x^a_{eq}$ to include the temperature gradient: 
\begin{align}
\begin{split}
x^a_u(\vR, \hp, \epsilon) =  
\left[1+ \gamma_{u}^R(\hp, \epsilon) \tilde \gamma_{u}^A(\hp, \epsilon)\right] \pder{\Phi_0}{T} \times
\\
\times \left[\delta T(\vR) - \ell_u(\hp,\epsilon)\hp\cdot\grad T \right]
\,.
\end{split}
\label{eq:xAU}
\end{align}
This expression with uniform gradient $\grad T = const$ satisfies \refE{eq:xAlin}. 
The $\grad T$ term in \refE{eq:xAU} is odd in momentum, and after angular integration 
in \refE{eq:sigmaAsc} it does not contribute  
to self-energy $\sigma^a$ in even-$\hat p$ superconductor. 
Consequently, only the first term, $\propto \delta T(\vR)$, determines the dynamical self-energy.  
In entirely uniform superconductor, $x^a$ and $\tilde x^a$ would be trivially related and 
result in local equilibrium self-energy
$ \sigma_u^a= (\sigma^R_{0,imp}-\sigma^A_{0,imp} ) (\partial\Phi_0/\partial T) \, \delta T(\vR)$. 
This is important since after substitution of this expression together 
with \refE{eq:xAU} into \refE{eq:xAlin} the arbitrary $\delta T(\vR)$ drops out. 
Non-uniform order parameter means different history for $x^a(\hat p)$ 
and $\tilde x^a(\hat p)$ along a trajectory and the self-energy 
$\sigma^a$ even in uniform part does not fully recover the local equilibrium 
dependence on $\delta T$.

On the other hand, by similar symmetry arguments, the heat current in the uniform part of the superconductor is 
independent of $\delta T$ and is completely 
determined by the $\grad T$ term of $x^a_u$. We use it to set the value of the 
temperature gradient from fixed $\vj_h^{\rm BC}$. 
Inserting \refE{eq:xAU} into \refE{eq:qLinResp}, and using 
$\gamma$-symmetries \refE{eq:symmRic}, we obtain uniform-state current 
\beq
\vj_h^{\rm BC}=- \kappa_u  \grad T \,,\qquad
\kappa_u=\int\limits_{-\infty}^{+\infty} d\epsilon\, \kappa_u(\epsilon) \,,
\label{eq:kapU}
\eeq
where the thermal conductivity has this Boltzmann-like representation
\beq
\kappa_u(\epsilon)=
\frac{  v_F \epsilon^2}{2 T^2 \cosh^2(\epsilon/2T)}  
  \fsav{\hp_x^2 N(\hp,\epsilon ) v(\hp,\epsilon) \bar \tau(\hp, \epsilon)  } \,. 
\label{eq:kapUepsilon}
\eeq
Here $N(\hp,\epsilon)$ is the density of states, 
$\bar \tau(\hp, \epsilon)=[\ell(\hp, \epsilon) + \tilde \ell(\hp, \epsilon)]/2 v_F$ is a
scattering time defined using relaxation length \refe{eq:elldef} 
(in Unitary or Born limits $\ell(\hp,\epsilon) = \tilde \ell(\hp,\epsilon) \equiv \ell(-\hp,-\epsilon)$). 
The group velocity for quasiparticles (QPs) with momentum $\hp$ and energy $\epsilon$ is given by 
\beq 
v(\epsilon,\hp) = v_F\frac{1-|\gamma_{u}^R(\hp, \epsilon)|^2}{1+|\gamma_{u}^R(\hp, \epsilon)|^2} \,. 
\eeq 
From this, the  velocity of  QPs in superconductor is always  smaller than  $v_F$; 
also in the clean limit one recovers the well-known result  
$v(\hp, \epsilon) = v_F \sqrt{\epsilon^2-\Delta^2(\hp)}/|\epsilon|$.
Typically, heat transport in uniform superconductors 
is analyzed as interplay between the 
density of states $N(\hp, \epsilon )$
and effective elastic mean free path 
\beq
\ell_e \equiv \bar\tau (\hp, \epsilon)v(\hp, \epsilon) \,,
\label{eq:ell_e}
\eeq
where the latter plays a more dominant role. 
This discussion is moved to Appendix \ref{app:scattering_ell}, 
with the main result presented in Figs.~\ref{fig:uniform} and \ref{fig:kUni} there. 
 
\begin{figure}[t]
 \includegraphics[width = 0.99\linewidth]{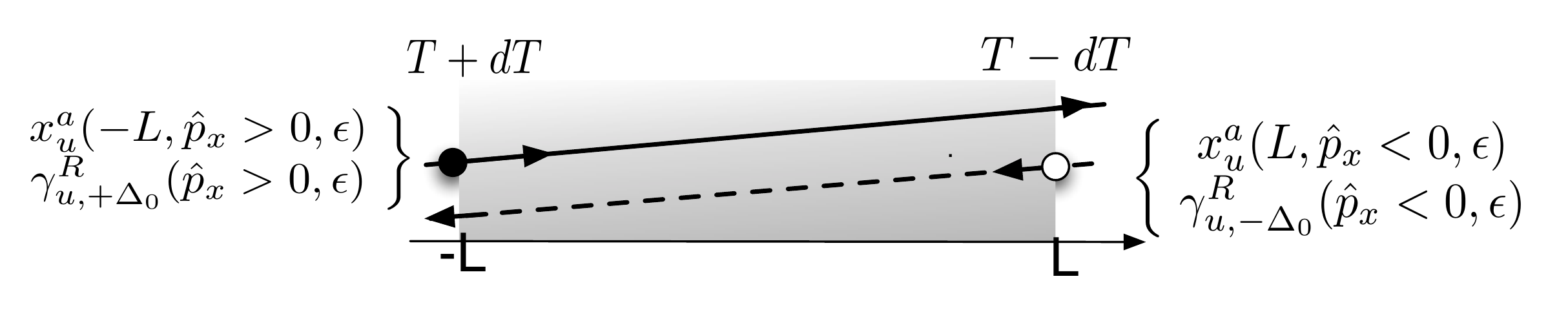}
\caption{ \label{fig:BCmodel} 
 Numerical integrations of  Eqs. \refe{eq:gamma} and \refe{eq:xAlin}, in the
shaded region, performed from $x=\mp L$ to $x=\pm L$ for right/left going ($\hp_x \lessgtr 0$) 
trajectories along  $\hp$. We start the numerical integration  at the white/black circles with the uniform Riccati amplitude given by  Eqs. \refe{bc:gamma} and \refe{bc:xa}, see main text.     The (half-)  temperature bias $dT$   is
the unknown that we numerically  determine to satisfy
\refE{bc:current}.  
}
\end{figure}
 
We are now ready to write the initial values of 
coherence and distribution functions for numerical integration 
Eqs. \refe{eq:gamma} and \refe{eq:xAlin} along quasiclassical trajectories. 
We start the integration well away from any domain walls, inside the uniform 
part of superconductor, at $x=\pm L$, Fig. \ref{fig:loc_glo}. 
The distance between the initial point on the trajectory 
and the first domain wall should be much greater than $\ell(\hat p, \epsilon)$, 
which might be difficult to satisfy for all energies and momenta, 
especially in clean superconductor. 

The equilibrium coherence functions at $x=\pm L$ arrive from infinity with the uniform bulk values, 
Fig.~\ref{fig:BCmodel}:
\begin{align}
\begin{split}
\gamma_0^R(-L, \hp_x>0, \epsilon) = \gamma^R_{u, 0}(\Delta(x=-\infty,\hp), \hp, \epsilon) \,, 
\\
\gamma_0^R(+L, \hp_x<0, \epsilon) = \gamma^R_{u, 0}(\Delta(x=+\infty,\hp), \hp, \epsilon) \,. 
\end{split}
\label{bc:gamma}
\end{align}
We will position the center of  domain walls symmetrically around $x=0$, ensuring equivalent 
temperature drops $dT$ on the left and right, to accelerate numerical integration. 
The initial values for the anomalous distribution $x^a(\pm L)$ is given by \refE{eq:xAU} 
with temperature gradient fixed by the heat current in uniform region: 

\begin{align}
\begin{split}
\grad T = - \frac{ j^{\rm BC}_h}{\kappa_u} \, \hat x  \,,
\\
x^a(- L, \hp_x>0, \epsilon) = (1+ \gamma_u^R\tilde \gamma_u^A) \pder{\Phi_0}{T} 
\left[dT + \ell \,\hp\cdot\grad T \right] \,,
  \\
x^a(L, \hp_x<0, \epsilon) = (1+ \gamma_u^R\tilde \gamma_u^A) \pder{\Phi_0}{T} 
\left[-dT + \ell \,\hp\cdot\grad T \right] \,.
\end{split} 
 \label{bc:xa} 
\end{align}
The unknown temperature drop $dT$ is determined, for a given $j_h$, through self-consistent 
calculation of anomalous self-energies $\sigma^a(x,\epsilon)$ at each $\epsilon$.  
Starting with some guess for $\sigma^a(x,\epsilon)$ we solve \refE{eq:xAlin} for $x^a(x, \hp, \epsilon)$ 
with boundary conditions \refe{bc:xa}. From distribution function we find 
$g^a(x, \hp, \epsilon)$, \refE{eq:RicKga}, and then obtain new values for $\sigma^a(x,\epsilon)$, \refE{eq:sigmaAsc}. 
This process is repeated until self-energy has converged. 
The linearity of all equations assures that all the parameters are linear combinations of $dT$ and $\nabla T$ terms:
\begin{align}
\begin{split}
x^a(x, \hp, \epsilon) = x^a_1(x, \hp, \epsilon) dT + x^a_2(x, \hp, \epsilon) \nabla T \,,  
\\
g^a(x, \hp, \epsilon) = g^a_1(x, \hp, \epsilon) dT + g^a_2(x, \hp, \epsilon) \nabla T  \,,
\\
\sigma^a (x, \epsilon) = \sigma^a_1(x, \epsilon) dT + \sigma^a_2(x, \epsilon) \nabla T \,, 
\end{split} 
\end{align}
and similarly the current,
$j_h(x) = \kappa_1 dT + \kappa_2 \nabla T  = const= \kappa_u \nabla T$, 
that is equal to the input current at the boundary. 
After self-consistent determination of the coefficients $\kappa_{1,2}$ through the above procedure 
we determine the temperature drop
$$ 
dT = \frac{ \kappa_u - \kappa_2}{\kappa_1} \nabla T  \,.
$$
In the uniform case one has $dT= L j_h^{BC}/\kappa_u$.

Here we also want to note that the presence of topological domain walls in the order 
parameter is reflected in features of the heat current arbitrarily far from the 
nonuniform region, and is indirectly encoded in the
choice \refe{bc:xa} for $x^a$ at the integration boundaries. 
For example, in a uniform superconductor the spectral current is given by $\kappa_u(\epsilon) |\grad T|$, 
whereas in nonuniform superconductor we have $j_h(\epsilon) \ne \kappa_u(\epsilon) |\grad T|$, 
which is obvious if we are right at the domain wall, and thus everywhere else due to 
the conservation of the spectral current, as shown in Appendix \ref{app:conservation}. 
To recover the heat current spectrum of the uniform state far from the nonuniform region, 
one requires presence of nonelastic collisions that are not included in the theory. 
By contrast, 
the local equilibrium picture includes the nonelastic collisions implicitly in the definition of 
the local temperature $T(x)$ in Fermi distribution.

\section{Heat flow across domain walls \label{sec:non-uni}} 
 
We  now apply the developed formalism to nonuniform $d$-wave superconductor, 
and investigate heat transport 
across an array of $N_\dw$   domains walls  equally spaced  with a period  $X_\fflo$ along the $\hat x$-axis.   
%
Each domain wall has a width of several coherence lengths that we 
define as 
$$
\xi = \frac{\hbar v_F}{2\pi k_B T_c}
$$
($T_c$ is the transition temperature of clean supercnductor).
The uniform heat current flows from left to right $\vj_h= j_h \hat x$, 
and we consider translationally invariant system along the  $ \hat y$-direction, 
so that all functions depend only on $x$-coordinate. 

For convenience, we now set a unit gradient at the boundaries 
$\grad T = - \hat x$ in \refE{bc:xa},
giving $j_h = \kappa_u\times 1$. 
%

Due to the factor $|\epsilon \,\partial_T\Phi_0| = \epsilon^2/[2T^2\cosh^2(\epsilon/2T)]$ 
the heat current is mainly determined by quasiparticles
with energies in the window $[T,5T]$. 
We introduce $\epsilon_T=2.5 \, T$ as a characteristic quasiparticle energy at a given temperature $T$.

\subsection{Single domain wall}
\label{sec:singleDW}

\begin{figure}[t]
\includegraphics[width = 0.7\linewidth]{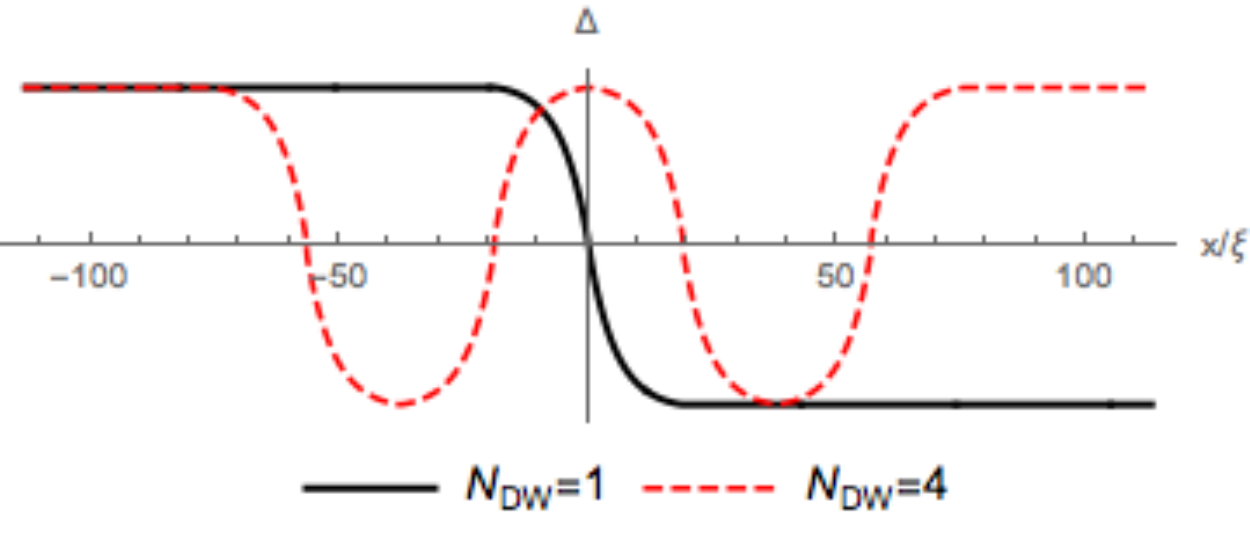}\\
\includegraphics[width = 0.7\linewidth]{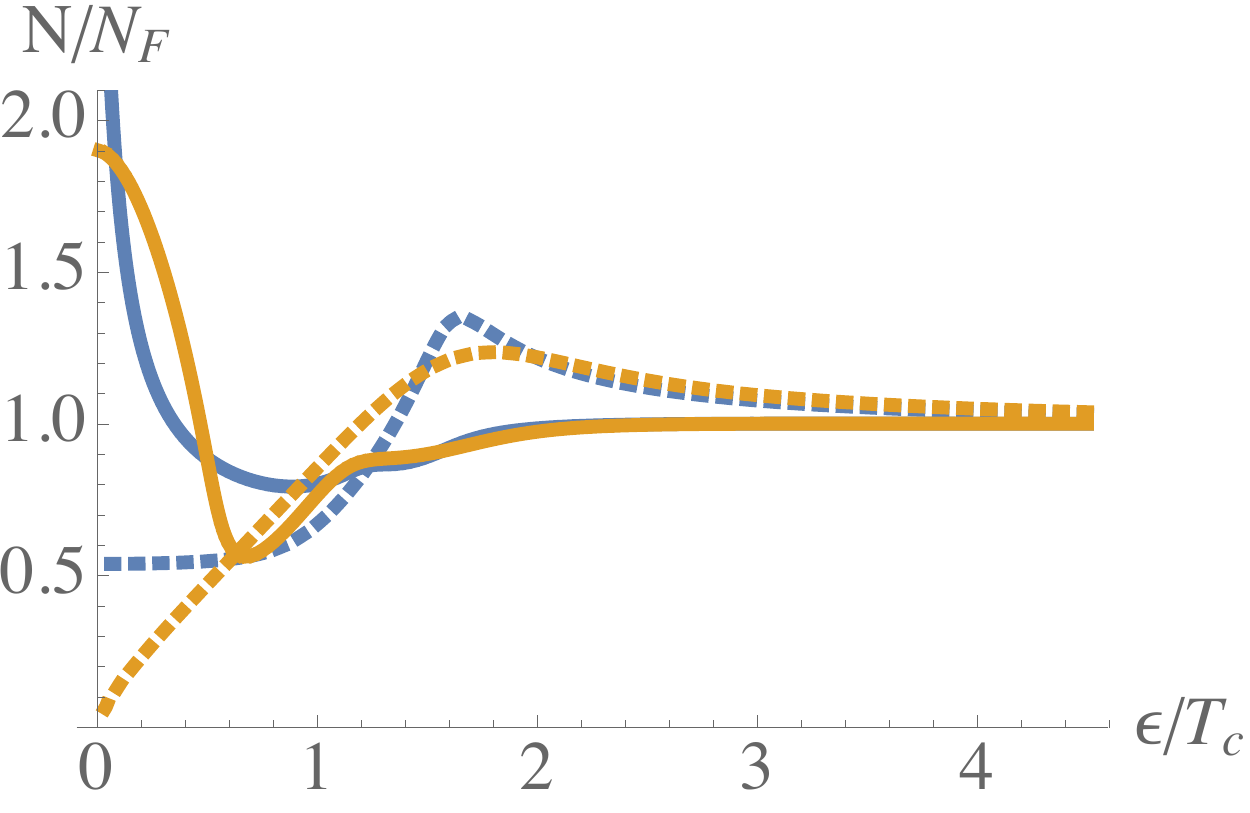}
\caption{ \label{fig:DWDoS} 
Upper panel: self-consistent OP profile $\Delta(x)$ for a single DW (solid line). 
This solution is used to construct a non-self-consistent profile with $N_{DW}$, 
case of 4 DWs with separation $X_\fflo\approx20\xi$ is shown by the dashed line.
Lower panel: local density of states (DoS) in Born (orange) and unitary (blue) limits 
for $ \ell_\No=   \pi \xi/0.3$. 
At the domain (solid lines), the peak at zero energy indicates the Andreev bound states (ABS). 
}
\end{figure}

We first look at the heat transport across a single domain wall (DW) centered at $x=0$ ($N_\dw=1$). 
The domain wall is enforced through the boundary condition $ \Delta_0(\pm L)= \pm \Delta_u$. 
It is self-consistently computed together with  the local impurity self-energy $\sigma _\imp(x, \epsilon)$
via  Eqs. \refe{eq:scDelta} and   \refe{eq:scImp}. 


With the domain wall centered at $x=0$, we use symmetry $\Delta(-x) = -\Delta(x)$ to speed up 
numerical calculations through relations: 
\begin{align}
\begin{split}
\gamma^R_0(x, \hp, \epsilon)=- \gamma_0^R(-x, -\hp, \epsilon),\\ 
\gamma^R_0(x, \hp, \epsilon)=- \tilde \gamma_0^R(x, -\hp, \epsilon),
\end{split}
\label{eq:sym}
\end{align}
and similar ones for self-energies, 
\begin{align}
\begin{split}
\sigma^R(x, \hp, \epsilon)=  \tau_z \sigma^R(-x, \hp, \epsilon)\tau_z,\\ 
\sigma^R(x, \hp, \epsilon)=  [ \sigma^R(x, \hp, \epsilon)]^{tr},\\ 
\sigma^a(x, \epsilon)= - \tau_z \sigma^a(-x,  \epsilon)\tau_z \,.
\end{split}
\end{align}

Technically we proceed as follows.  First, we  obtain the
order parameter profile $\Delta_0(x)$, shown in Fig. \ref{fig:DWDoS}(a), using Matsubara technique. 
With the known mean field profile, we integrate \refE{eq:gamma} for real energies to determine the
equilibrium values of $\gamma_0^R(x, \hp, \epsilon)$ and impurity $\sigma^{R/A}_{\rm 0,imp}(x,\epsilon)$.  
They are then used  as input parameters in equation \refe{eq:xAlin} for the 
anomalous amplitude $x^a$. 
The last step is the self-consistent calculation of the temperature drop $dT$ together with anomalous 
self-energy $\sigma^a$. 


\begin{figure}[t!]

\includegraphics[width = 0.9\linewidth]{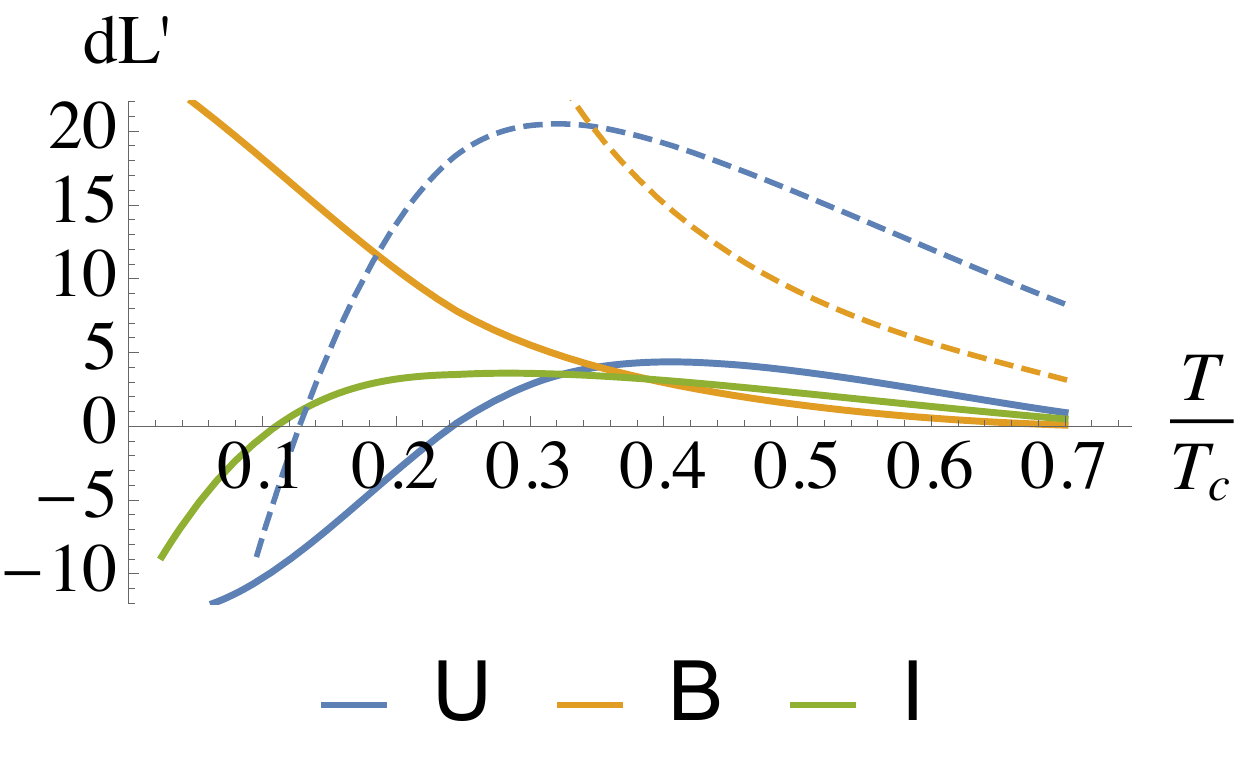}
\caption{ \label{fig:dT}  
Effective change in thermal length $dL'$ (in units of $\xi$) across a single domain wall relative to the uniform case, 
as function of temperature. 
Numerical system size is $2L=16\pi \xi$.  
Different colors correspond to unitary limit (U, blue), 
Born (B, orange) and intermediate phase shift $\delta = \pi/4$ (I, green). 
Solid lines are for scattering rate $2 \Gamma \sin^2\delta = 1/\tau_N = 0.6 T_c$, 
dashed lines are for a cleaner case 
$1/\tau_N = 0.2 T_c$. 
In the unitary limit, $dL'$ is non-monotonous, and at low temperature $ T < W_{\rm imp}$,  the heat conductance through 
a domain wall is larger than in uniform case. 
}
\end{figure}

We compare the  temperature drop $dT$ with the drop $dT_u=(j_h/\kappa_u)L = |\grad T|_{u} L$ 
that would appear if the superconductor was uniform. 
Then $dT > dT_u$ 
corresponds to a suppression of   ability to transport heat across domain walls, 
while $dT-dT_u < 0$ represents an  enhancement of heat conductivity. 
The numerical results for    transport     across the domain wall  are presented in Fig. \ref{fig:dT}, 
where we plot the temperature drop across a domain wall for a given heat current, relative to the 
uniform configuration. We define the parameter $dL'$ with dimension of length
$$
dL'= \frac{dT-dT_u}{j_h/\kappa_u}  
\,,
$$
that can be interpreted as effective ``thermal length'' of 
the domain wall, in units of coherence length $\xi$. 

At high temperatures the behavior of the thermal transport is the same for all impurities, 
with a loss of effectiveness in energy transport. At low temperatures, however, the 
behavior is remarkably different in Born and Unitary limits. For weak impurity scattering potential 
the domain wall presents a barrier for heat transport resulting in a larger temperature drop required to 
maintain current $j_h$. The strong scatterers have the opposite effect - the heat current flows through a 
domain wall more efficiently than in the uniform case. 

The origin of such peculiar behavior is in the interplay between two effects of 
the Andreev bound states at the domain wall: the change of spectrum and the hybridization of 
the bound states with the impurity band states. 

\begin{figure}[t!]

   \includegraphics[width = 0.85\linewidth]{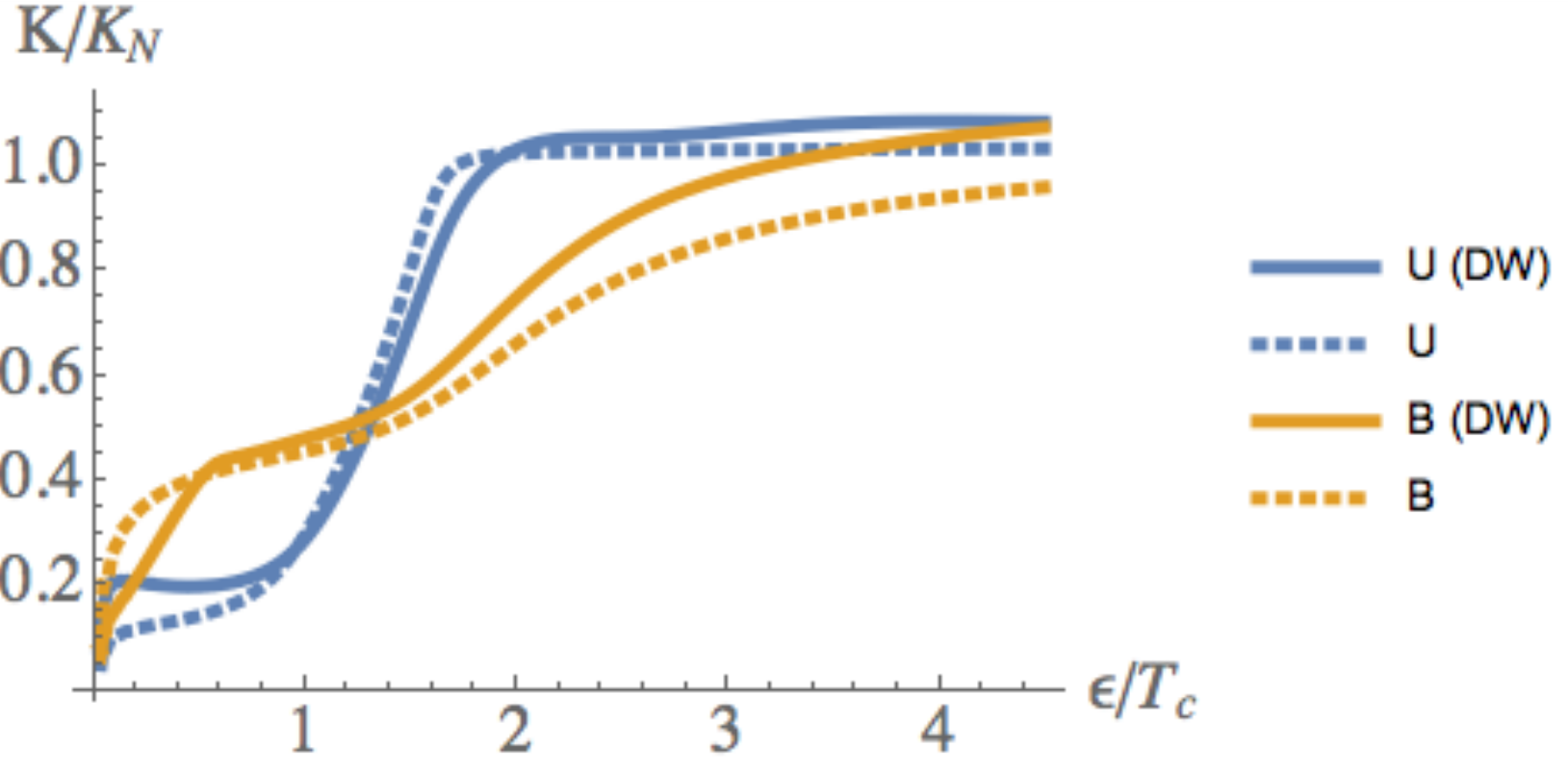}
\caption{ \label{fig:kernel_imp}  
Energy dependence of the heat current kernel at $T = 0.3 T_c$ for transport across a domain wall
(solid lines) and the uniform superconductor (dashed lines) for $ \ell_\No=  \pi \xi/0.3$.  
In Born or clean limit (orange lines) the ability to transport heat at low energy is suppressed  by the presence of a domain: at
low energy $K_\dw(\epsilon)<K_u(\epsilon)$. By contrast, in unitary limit (blue lines) the coupling between 
impurity band and Andreev bound states enhances energy transport: at  low energy $K_\dw(\epsilon)>K_u(\epsilon)$.  
} 
\end{figure}

The spectral effect is a result of Andreev bound states `stealing' spectral weight 
from continuum quasiparticles states above the energy gap. 
In bulk, the only available quasiparticles with $\epsilon > |\Delta(\hp)|$ participate in the energy transport. 
As these quasiparticles enter the domain wall region with fewer available states they experience 
Andreev reflection that leads to suppression of the heat conductivity. 
This effect can be quantified by looking at a clean superconductor. 
In this case equation for the distribution function \refe{eq:xAlin} has no impurity-generated right-hand-side, 
and the relaxation length (mean free path) 
$1/\ell_\Delta = 2\, \im[ \gamma^R \tilde\Delta]/v_F$ 
is determined purely by the density-of-states effects. 
Details of this analysis are presented in Appendix \ref{app:clean}. 
Effects of the spectral weight reduction and Andreev reflection processes 
appear in 
the heat current kernel $K(\epsilon, \hp)=j_h(\hp, \epsilon)/( \epsilon\, \partial_T \Phi_0)$, shown in
Fig.~\ref{fig:kernel_imp}, 
at energies $\epsilon \sim \Delta$ 
and play the most important role at higher temperatures. 

\begin{figure}[t]
   \includegraphics[width = 0.85\linewidth]{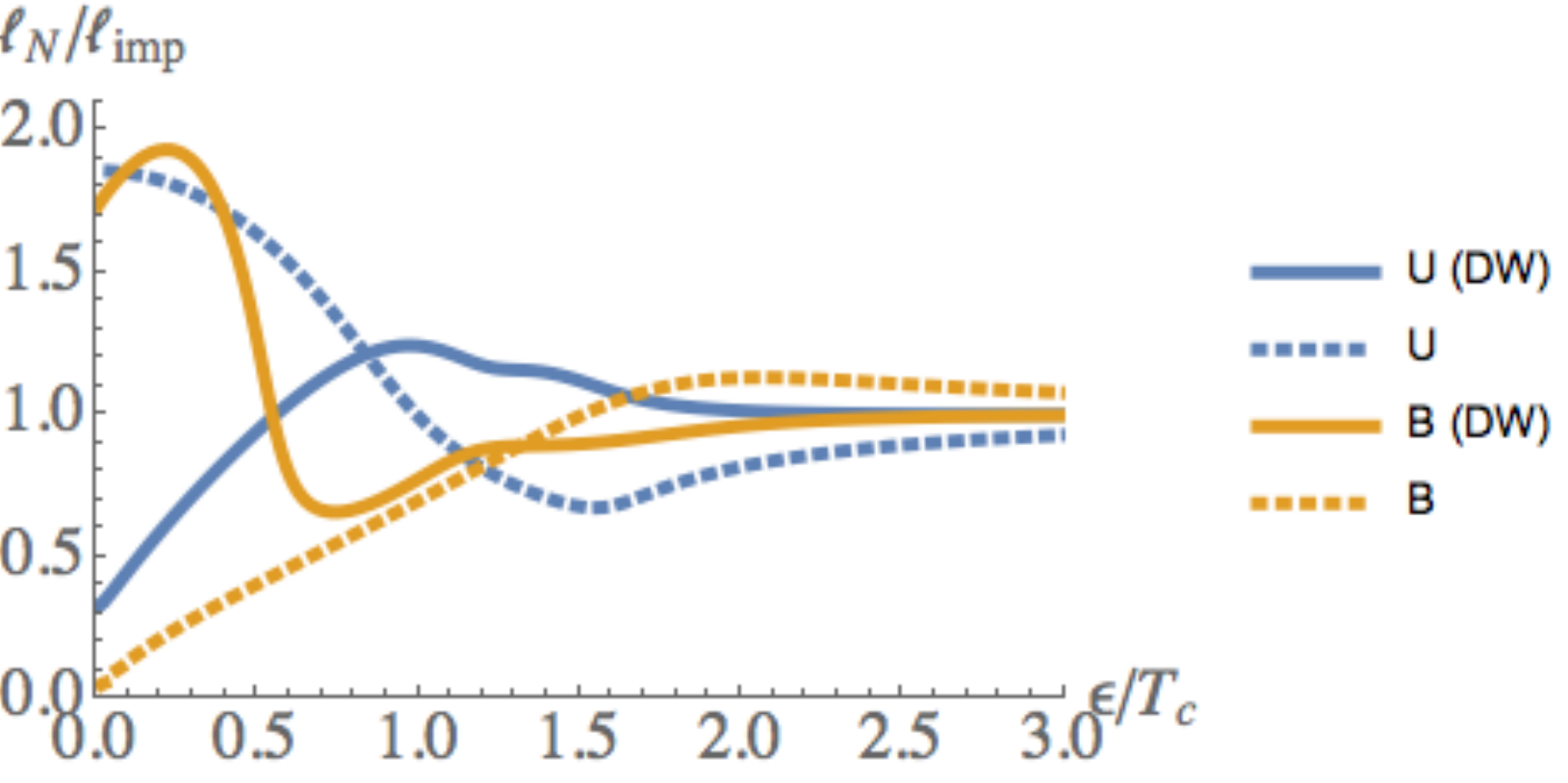}
\caption{ \label{fig:l_imp}  
Inverse local impurity scattering length 
$\ell_N/\ell_{\rm imp}(x,\hp,\epsilon) \approx\ell_N/ \ell_{\rm imp}(\epsilon, x)$ (weakly dependent on momentum directions), 
as a function of energy, for $ \ell_\No=  \pi \xi/0.3$. 
In Born limit (orange),  the mean free path is large in the bulk (dashed lines) and becomes small 
at the domain wall (solid lines). 
For unitary scattering (blue), on the right, this behavior is reversed: the zero-energy peak in the DOS 
results in suppression of scattering rate at the domain wall and longer mean free path. 
}
\end{figure}

At very low temperatures $T \ll T_c$, the interaction of low-energy bound states with impurities comes out to  
the front stage, while we find that $\ell_\Delta$ is only slightly modified by impurities. 
The impurity scattering effects appear in \refE{eq:xAlin} through anomalous self-energy and 
local scattering length 
$1/\ell_\imp = 2\, \im[ \gamma_0^R \tilde\Delta_\imp^R-\Sigma^R]/v_F$. 
This length is positive and finite, depends on directions very weakly and can be approximated by 
$\ell_\imp(x, \epsilon) \approx \fsav{\ell_\imp(x, \hp, \epsilon)}$. 
The impurity scattering creates a band of mid-gap states, 
which hybridize with Andreev bound states. 
Such hybridization depends strongly on the strength of the impurities and may lead 
to a significant `renormalization' of scattering features in the vicinity of the domain wall, 
as shown in Fig. \ref{fig:l_imp}.      
In the unitary limit, Andreev states' interaction with impurity band leads to suppression of scattering 
and long lifetime of close-to-zero-energy quasiparticles. This results in an effective `wormhole' across the
domain wall region for these quasiparticles, and an enhancement of heat conductivity at low temperature, see Fig.~\ref{fig:kernel_imp}. 
In the Born scattering limit, on the other hand, the impurity band is weak,
and its presence cannot compensate Andreev reflections. In this case, for all temperatures, 
the heat transport is suppressed across the domain wall.

\subsection{Multiple  domain walls\label{sec:DWs}  }

To model the periodic structures of FFLO states we investigate 
transport across a set of domain walls. 
Since the main effects come from the density of states and scattering, we omit the self-consistent calculation 
of the order parameter, and simpy `build' a lattice of 
$N_\dw$ equally spaced domains with an arbitrary period $X_\fflo$, 
taking the single domain profile as a unit cell, as shown in Fig.~\ref{fig:DWDoS} for $N_\dw=4$. 
We place the domains symmetrically around $x=0$ and use this symmetry to reduce computation time. 

 
 \begin{figure}[t] 
\includegraphics[width =0.85\linewidth]{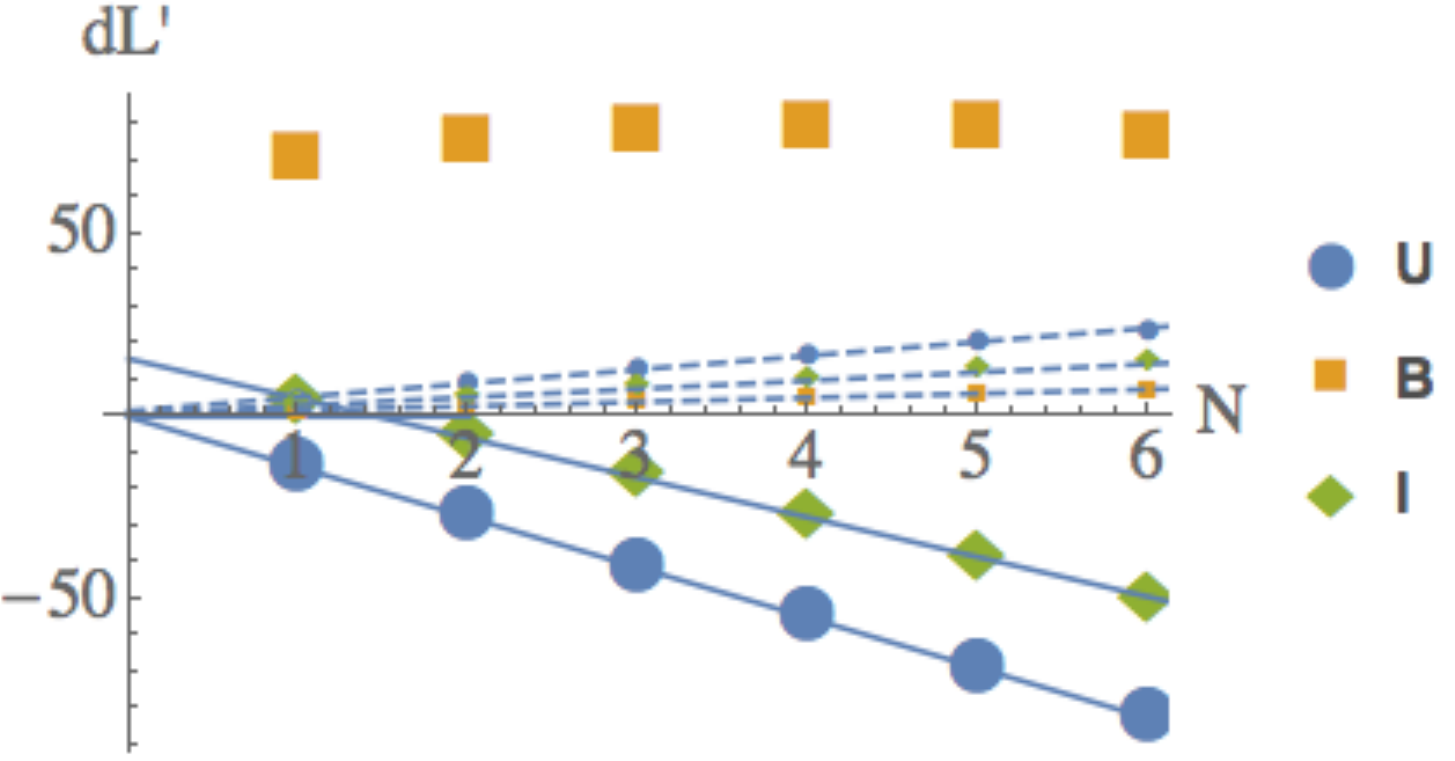}
\caption{\label{fig:dTDWs1} 
Effective thermal length $dL'$ (normalized by $  \xi$ ) across $\Ndw$ domain walls, 
for low temperature $T/T_c= 0.05$ (large symbols, solid lines)
and intermediate temperature $T/T_c=0.5$ (small symbols, dashed lines). 
The scattering rate $1/\tau_N=0.3T_c$ is used for various impurity strengths: 
Born (B), Unitary (U) and intermediate $\delta=\pi/4$ (I). 
This is an `independent domain walls' regime where the heat conductivity contributions 
from each domain add up, as is clear from linear dependency 
$c_1\Ndw+c_2$ shown by lines. 
At low temperature the unitary and intermediate strength disorder has negative slope 
consistent with single-domain result in figure \ref{fig:dT}, coming from low-energy states' transport.  
At intermediate temperature we have a suppression of heat flow due to independent Andreev reflection 
processes, with positive slope and linear increase in the thermal length  $dL'$ with $\Ndw$. 
In Born limit at low temperature the dependence is more complicated due to 
large extent of bound states and more intricate impurity band energy dependence for $T \sim W_\imp$. 
}
\end{figure}

There are several effects that influence the transport across multiple domain walls. 
First one is the trivial (incoherent) accumulation of effects from all domains that are 
independent in this case. 
This happens when the mean free path, \refE{eq:ell_e}, 
is shorter than the spacing $X_\fflo$ between domain walls, and the spatial extent of the bound 
states also exceeds this length, 
$X_{\rm ABS}[\hp] \approx  v_F/\sqrt{\Delta^2(\hp)+ W^2_\imp} \gg X_\fflo$, 
where $ W_{\rm imp}$ is the impurity bandwidth. 
Independent domain walls lead to linear dependence of the heat conductivity on the number of domain walls $N_\dw$, 
based on the temperature regime and single-domain result as in figure \ref{fig:dT}. 
Such behavior is expected for reasonably dirty superconductors. 
Full numerical results for domain wall spacing $X_\fflo\approx 18\, \xi$ are shown in 
Fig.~\ref{fig:dTDWs1} and in independent-domain  regime  are fitted with straight lines. 

When the superconductor is in the clean limit, and the domain walls are tightly spaced with $X_\fflo<X_\abs$,
the bound states belonging to neighboring domains can overlap, hybridize, and build up a
conduction band (hybrid transport). This is expected in FFLO phase when the order parameter is small and 
harmonic-like, with periods $\sim 5-10\xi$ 
rather than a combination of fully formed domain walls, or when the transport is dominated by the 
nodal quasiparticles since ABS states  can extend 
far beyond the DW region, especially in Born limit with tiny $W_{\rm imp}$. 

\begin{figure}[t] \includegraphics[width = 0.95\lw]{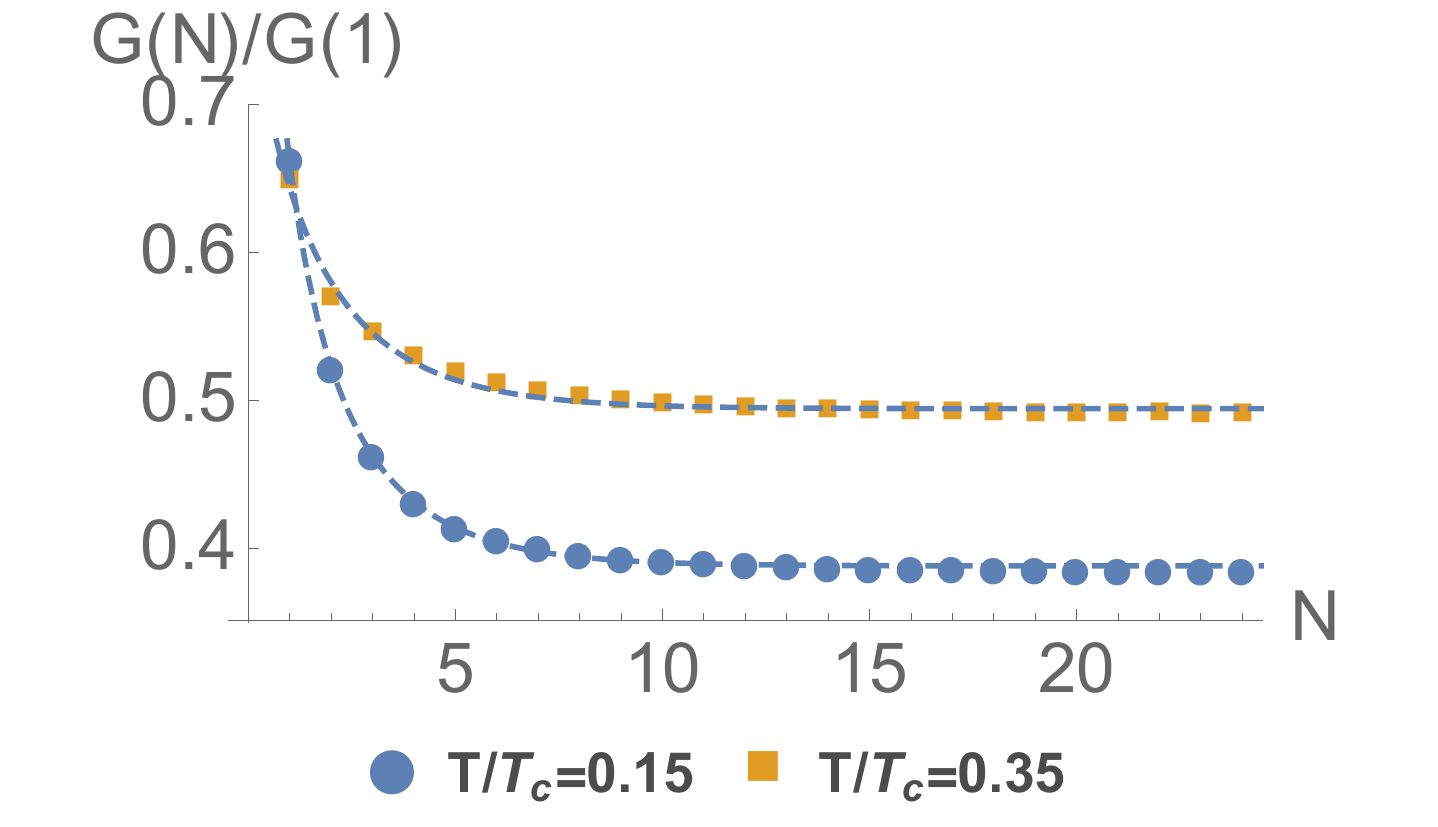}

\caption{ \label{fig:cleanDWs}  
Effect on thermal conductance of Andreev reflections from a set of $\Ndw$ domain walls (clean limit). 
Heat transport through more than ten consequitive domains is dominated by extended 
bound states along nodal directions on the Fermi surface. 
Phase space of those states and their contribution to the heat transport grow with temperature. 
The fitting line through numerical points is explained in the text. 
}
\end{figure}

If the spacing between the domain walls is somewhat longer, then the hybridization of bound states from different domains 
depends on their quasiclassical trajectory. 
In the anti-nodal direction, the ABS spatial extent is smaller than
$X_\fflo$ and ABS are spatially separated. Each domain is  the center of an Andreev reflection process. Consecutive
reflections add up and yield a power law reduction of the transmission of anti-nodal quasiparticles.
By contrast, in the nodal direction, the ABS extent is large. ABS at consecutive domains overlap and 
the transmission is rather insensitive to the number of domains. 
Together, they result in $\Ndw$-dependence seen in Fig.~\ref{fig:cleanDWs}. 
The heat conductance can be roughly fitted by a sum of nodal and anti-nodal contribution: 
$ g_{\rm  n.} +  g_{\rm  a.n.} t^\Ndw$, where conductance contribution from nodal quasiparticles 
$ g_{\rm  n.}$ grows with temperature, and transmission coefficient $t$ is only weakly temperature-independent. 


\subsection{Zeeman field \label{sec:zeeman}}

In this section we present the effects of a Zeeman field on heat transport across the nonuniform state,  
since the FFLO state is a result of competition between magnetization and 
condensation energies. Again, the main effect, we assume, is coming from the 
modification of the density of quasiparticle states that are shifted in energy 
by $\pm \mu H$ for up/down spins. 
We neglect the order parameter suppression due to magnetic field, 
which is relatively small at low temperature.\cite{VorontsovFFLO} 
Then spin up and spin down QPs are independent, and their contributions  to thermal transport add up.  

The dependence on spin enters equations 
\refe{eq:qLinResp},
\refe{eq:xAlin}, and boundary conditions 
\refe{eq:xAU} and
\refe{bc:xa} 
through energy shift in coherence functions $\gamma_0^{\rm R/A}(\epsilon \pm \mu H)$. 
The quasiparticle distribution 
function prefactor $\epsilon \partial_T \Phi_0(\epsilon, T)$ is not changed. 
We can use it to write the heat current as some spin-dependent kernel times the distribution 
function,
\beq
j_h = \sum_{s=\pm 1} \int d\epsilon \, K_s(\epsilon) \; \epsilon \partial_T \Phi_0(\epsilon, T) \,.
\eeq
We then can re-use the zero-field results to compute the thermal current including
the Zeeman splitting. In the Zeeman field the spin dependent kernel is simply the spin-independent kernel shifted energy: 
$K_s(\epsilon) = K(\epsilon - s \mu H)$.  
We then can transfer the  dependence on spins into the distribution function, without recalculating the kernel: 
\beq\label{eq:zeeman}
j^H_h= \frac12 \sum_\pm \int d\epsilon \, {j_h^0(\epsilon)} 
\left[ \frac{(\epsilon\pm \mu H) \partial_T \Phi_0(\epsilon\pm \mu H, T)}{\epsilon \partial_T \Phi_0(\epsilon, T)}\right],
\eeq
where $j^0_h(\epsilon)$ is the spectral heat current in the absence of Zeeman field. As a reminder,  Fig.~\ref{fig:kernel_imp} highlights the effect of impurity  on the kernel of heat current in absence of Zeeman field.   

\begin{figure}[t]
  \includegraphics[width = 0.8\linewidth]{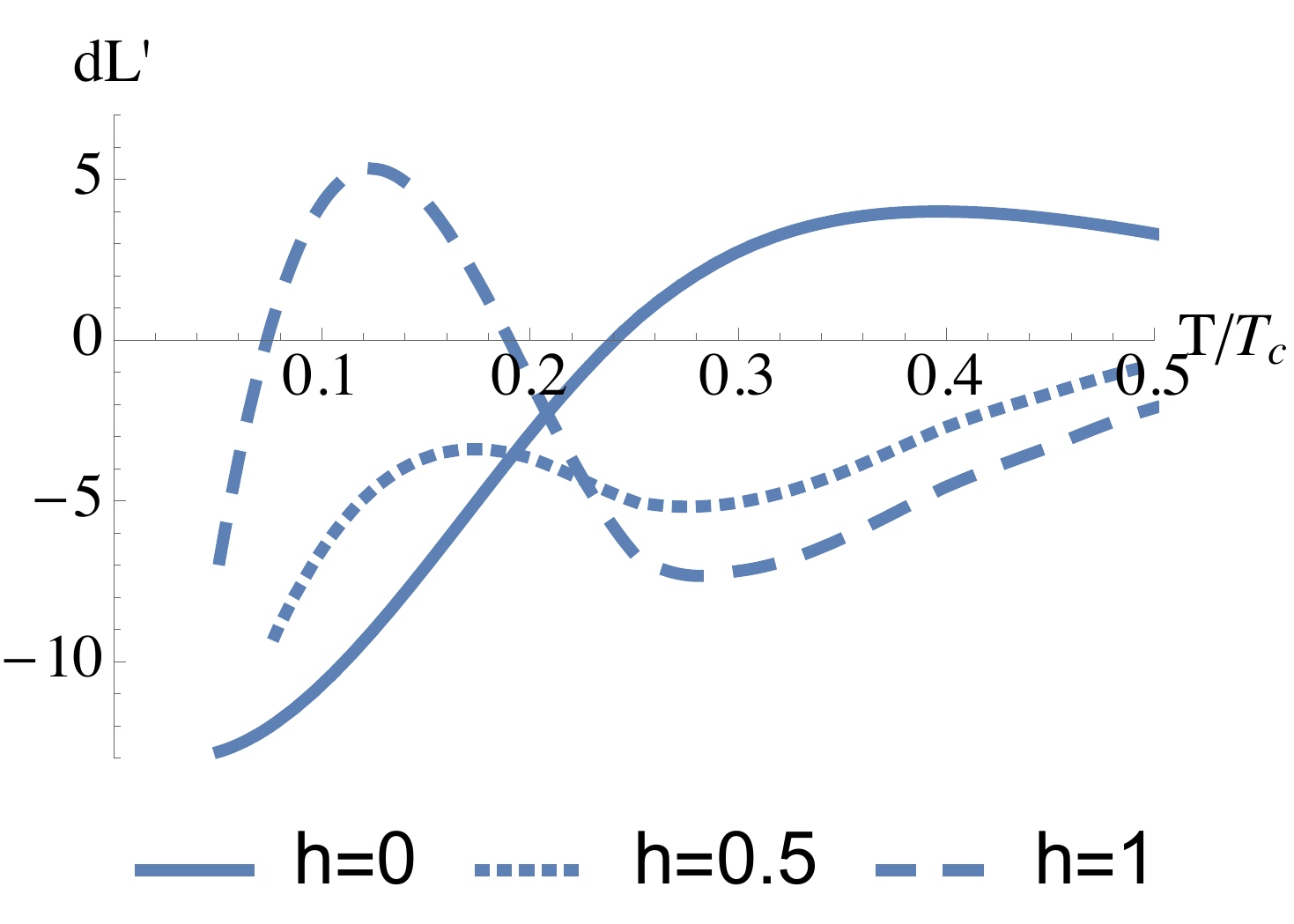}
\caption{ \label{fig:zeeman} 
Effect of the Zeeman field splitting $h = \mu H/T_c$ on thermal transport 
across a single domain wall. 
Unitary limit with scattering rate $1/\tau_N= 0.3\,T_c$. 
The bound states, shifted by $h=0.5$ contribute to a reduction of the thermal length, $dL'$,  in a wide range of temperatures. 
When the Zeeman shift is very large $h=1$ the contributions from bound and continuum states mix up leading to 
very non-monotonic temperature dependence. 
}
\end{figure}

The effect of Zeeman splitting of the states on thermal conductivity across a single domain wall 
is shown in Fig. \ref{fig:zeeman} for strong impurities. The bound states contribute most to the 
low-energy heat current and lead to increase in conductivity at low temperatures $T \lesssim \Delta \epsilon_{BS}/2.5$. 
From the $h=0$ curve the half-width of the bound states can be estimated as $\Delta \epsilon_{BS} \sim 0.4 T_c$.  
When the Zeeman field shifts the bound states by $ h = \mu H/T_c = 0.5 $ they dominate the heat transport in 
a wide range of temperatures leading to negative $dT-dT_u$. 
For even higher fields $h =1$, close to the critical field, the contributions of bound states with one spin projection 
mix with the continuum contribution with the other spin projection, leading to a non-monotonic 
temperature dependence of the heat conductivity. 
 


\section{Conclusions}

In this paper we have developed theoretical framework to investigate thermal transport 
in nonuniform superconductors. Our approach is based on fully self-consistent 
non-equilibrium quasiclassical Eilenberger-Keldysh technique, that takes into account, 
on the same footing, combined effects of impurity scattering, 
spatial variations of the order parameter and density of states, 
and the presence of Andreev bound states in strongly inhomogeneous environments. 

We applied this theory to compute the  thermal current across a periodic modulations of the 
order parameter, and domain walls, in a superconductor with $d$-wave pairing. 
Here we outline the key effects that govern transport in such systems compared with the 
uniform superconductors. 
First, Andreev bound states `trap' quasiparticles and cause a depletion of the 
continuum ($\epsilon > \Delta$) states near the domain wall, leading to Andreev 
reflection processes with particle-hole conversions. 
This results in  a reduction of heat transport across the domain wall, and this mechanism is dominant at 
intermediate temperatures and in clean superconductors. 
Another effect becomes relevant at low temperatures when disorder is present.  
Then the bound states at the domain wall interact with the low-energy impurity band.  
The coupling of the impurity band to localized Andreev states strongly depends on the type of impurity scattering. 
In Born limit this coupling increases scattering rate, while in unitary limit the scattering of low-energy 
quasiparticles is suppressed. These states have longer mean free path in the domain wall region 
resulting in an effective `wormhole' through the domain wall. At low temperature, below the width of  
the impurity band, transport is dominated by these states and with unitary impurities heat conductivity 
across the domain wall is higher than conductivity in the uniform state. 
This results in a very distinct non-monotonic feature of heat conductivity as a function of temperature, 
as one crosses from high- into low-energy regime. 
In a Zeeman field the difference between thermal transport in uniform and nonuniform phases is 
softened, but due to the opposite shifts of the up/down spin states, one can observe 
additional features in $T$-dependence of the heat conductivity, and non-monotonic $T$-dependence 
appears even in the Born limit. 
A grid of multiple domain walls generally amplifies transport properties of a single domain,
but in the clean limit one has to consider multiple-wall Andreev backscattering processes. 

These results show that thermal transport can be a 
useful probe to detect and study nonuniform states, such as Fulde-Ferrell-Larkin-Ovchinnikov phase that so far 
has been only identified using NMR technique.\cite{Mayaffre:2014dj} 
The approach that we developed will pave the way for future theoretical 
studies of heat transport near surfaces of superconductors with non-trivial 
surface states, in vortex lattices including vortex core states 
or for complete analysis of FFLO-type order parameter periodic structures.

\section{Acknowledgements}
This work has been done with NSF support through grant DMR-0954342.

\appendix
 
\section{Uniformity of currents 
\label{app:conservation} 
}

In the absence of inelastic scattering processes, the self-consistent solution of the Elenberger transport 
\refE{eq:Eil} together with impurity self-energies \refe{eq:scImp} 
guarantees uniform heat flow, and non-accumulation of heat, $\grad\cdot \vj_h = - \partial_t Q = 0$,
even in the presence of spatially-varying order parameter. 
The heat current is given by \refE{eq:heatqdef} which we repeat here:
\beq
\vj_h( \vR)=  2 N_Fv_F\int\limits_{-\infty}^{+\infty} \frac{d\epsilon}{4\pi i} \int d\hp\; 
[\epsilon \, \hp] \; 
\frac14 \tr\left\{g^K(\vR,\hp,\epsilon) \right\} \,.
\eeq
With only energy-conserving impurity collisions, all $\epsilon$ are independent, 
and we can consider divergence of the heat current kernel for single energy, 
$ \grad \cdot \vj_h(\vR,\epsilon) \propto \fsav{ \hp\cdot  \tr{\grad g^K} }$. 
Using equation for Keldysh component of \refE{eq:Eil} 
\beq
iv_F \hat{p} \grad g^K = g^K (\epsilon\tau_z - \sigma^A) - (\epsilon\tau_z - \sigma^R) g^K  +\sigma^K g^A - g^R \sigma^K
\eeq
we can split off the mean field self-energy $\Delta(\vR,\hat p)$, common for both retarded and advanced functions and zero
for Keldysh component, 
from the impurity self-energy. This allows us to write 
\beqa
&& \fsav{ \hp\,  \tr{\grad\cdot g^K} }   \propto 
 - \fsav{ \tr \{ [\epsilon \tau_z - \Delta, g^K] \} } +
\nonumber \\
&& 
\tr\left\{ \sigma_\imp^R\fsav{g^K}- \fsav{g^K}\sigma_\imp^A+ \sigma_\imp^K\fsav{g^A}-\fsav{g^R} \sigma_\imp^K  \right\}
 \nonumber
\\
&& = -0 + 0
\eeqa
where the first term is zero due to the traceless property of a commutator
and the second zero follows from the self-consistent relations between 
impurity self-energies and the Fermi-surface averaged propagators, \refE{eq:scImp}. 

Note that the order parameter self-consistency was not used in the above argument. 
It is however needed to conserve the charge/particle number. The formula for the particle current, written in terms of 
4-trace, acquires an extra $\tau_z$ matrix 
(and absence of $\epsilon$ factor):
\beq
\vj_e( \vR)=  2 N_Fv_F\int\limits_{-\infty}^{+\infty} \frac{d\epsilon}{4\pi i} \int d\hp\; 
\hp \; 
\frac14 \tr\left\{ \tau_z g^K(\vR,\hp,\epsilon) \right\} \,.
\eeq
Following the same line of arguments as for the heat current above, 
we notice that the impurity self-energy part vanishes due to same self-consistency 
as before but the commutator term with the mean-field order parameter is 
\beq
\int d\epsilon \fsav{ \tr \; \{ \tau_z [\epsilon \tau_z - \Delta \,,\, g^K] \} } 
= 2 \int d\epsilon \fsav{ \tr \;\left\{ \Delta \tau_z  g^K \right\} } \,,
\eeq
- vanishes if one uses the self-consistency on $\Delta(\vR,\hp)$  \refE{eq:scDelta}, 
ensuring non-accumulation of charge.

\section{Relative importance of density of states and mean free path.
\label{app:scattering_ell} 
}

To relate our results and treatment to previous work, in this appendix we present 
results for a uniform d-wave superconductor. 
The heat transport in a typical Boltzmann picture depends on a product of 
the density of states $N(\hp, \epsilon )$
and effective elastic mean free path 
\beq
\ell_e \equiv \bar\tau (\hp, \epsilon)v(\hp, \epsilon) \,,
\label{app:ell_e}
\eeq
The low-energy spectrum of a d-wave superconductor  
is strongly modified by the scattering of quasiparticles on impurities due to
the anisotropy of the order parameter structure.
Scattering on impurities results in formation of midgap states.\cite{Balatsky2006imprev}   
These impurity-bound states are 
extended in space and form a conduction `impurity' band with energy width $W_\imp$.
\cite{Graf1996,DurstLee2000}
This bandwidth is tiny in the
Born limit,  $W^B_\imp\approx 4 \Delta_0 \exp(-\frac{\pi\Delta_0}\Gamma)$, but can
be large in the Unitary limit where  $W^U_\imp\approx\sqrt{\pi \Delta_0\Gamma/2}$.  
  
\begin{figure}[t!]
 
  \includegraphics[width = 0.45\linewidth]{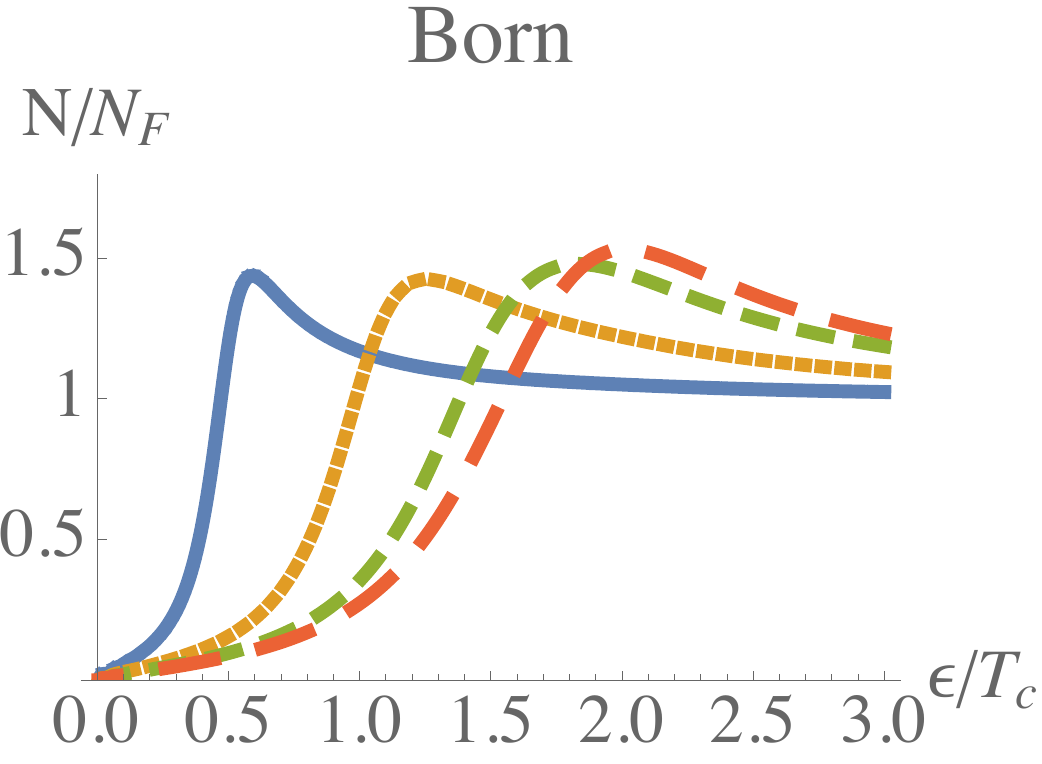} \llap{ {\includegraphics[height=0.2
\lw]{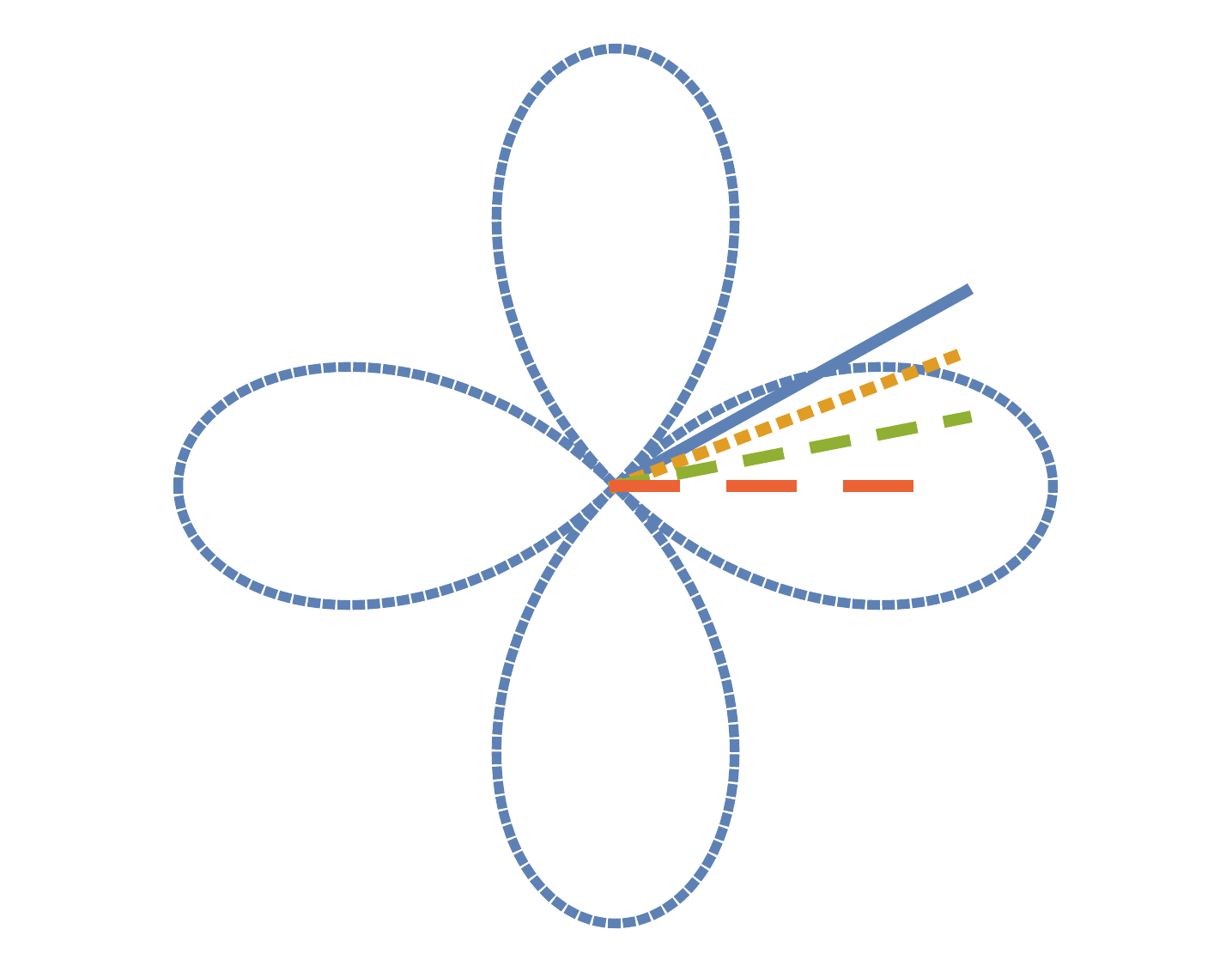}}} 
\includegraphics[width = 0.45\linewidth]{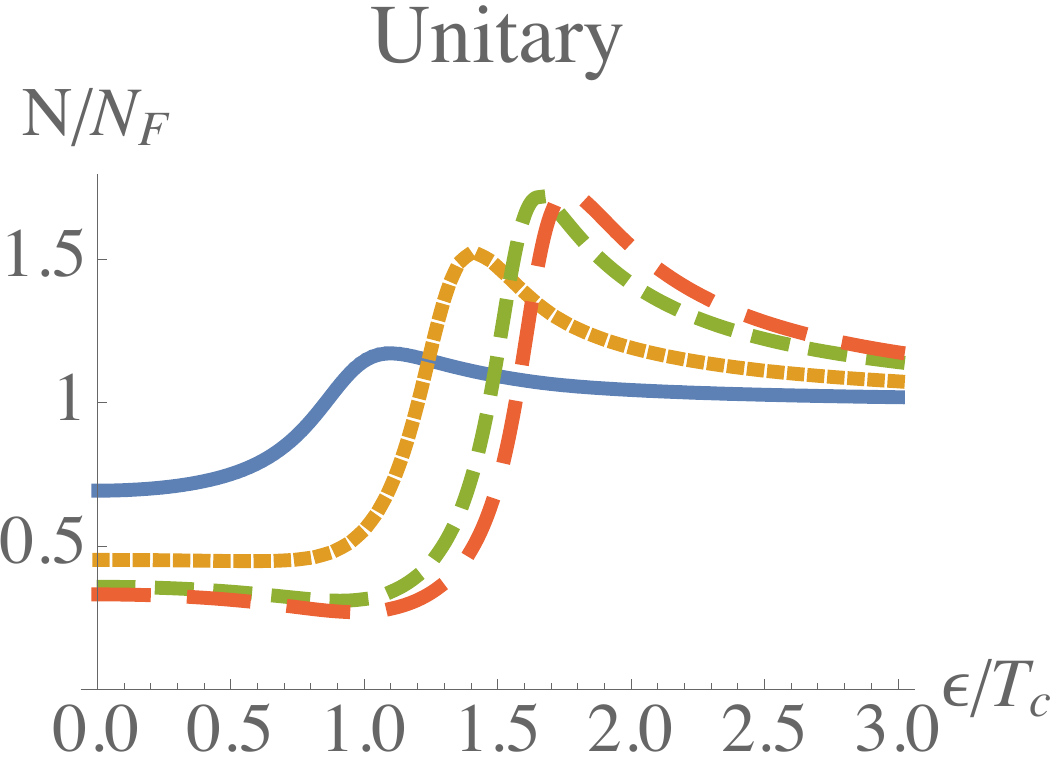}
 
   \includegraphics[width = 0.45\linewidth]{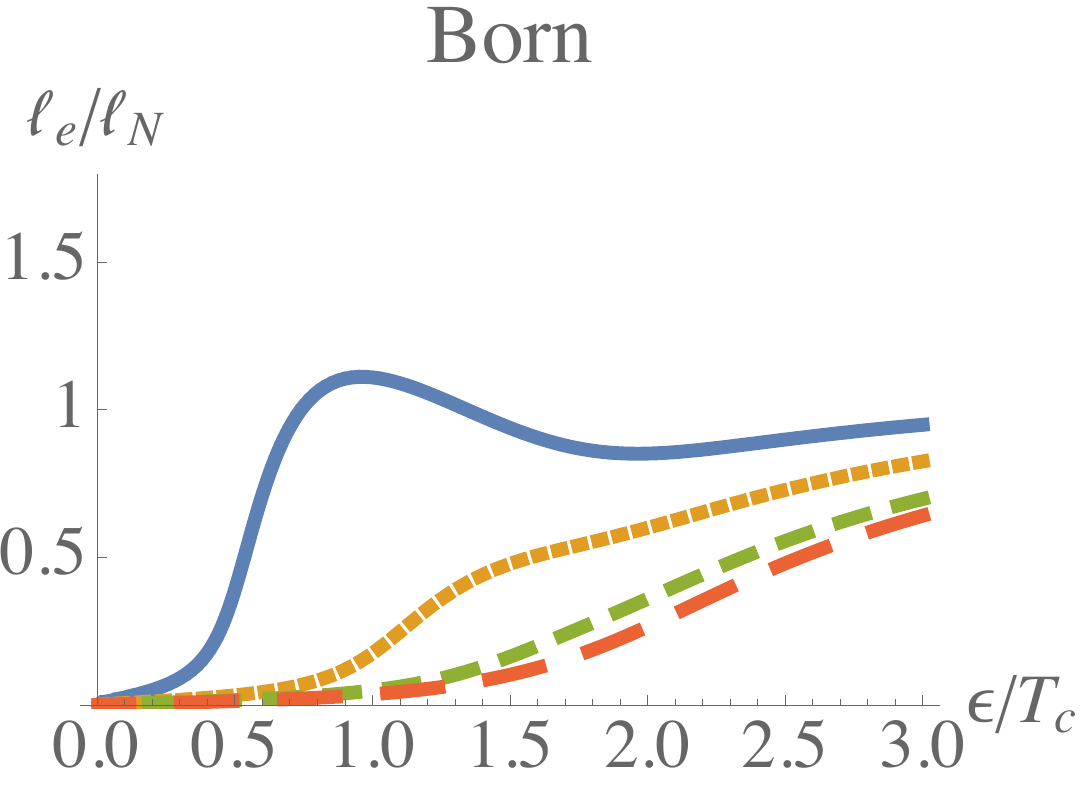} 
\includegraphics[width = 0.45\linewidth]{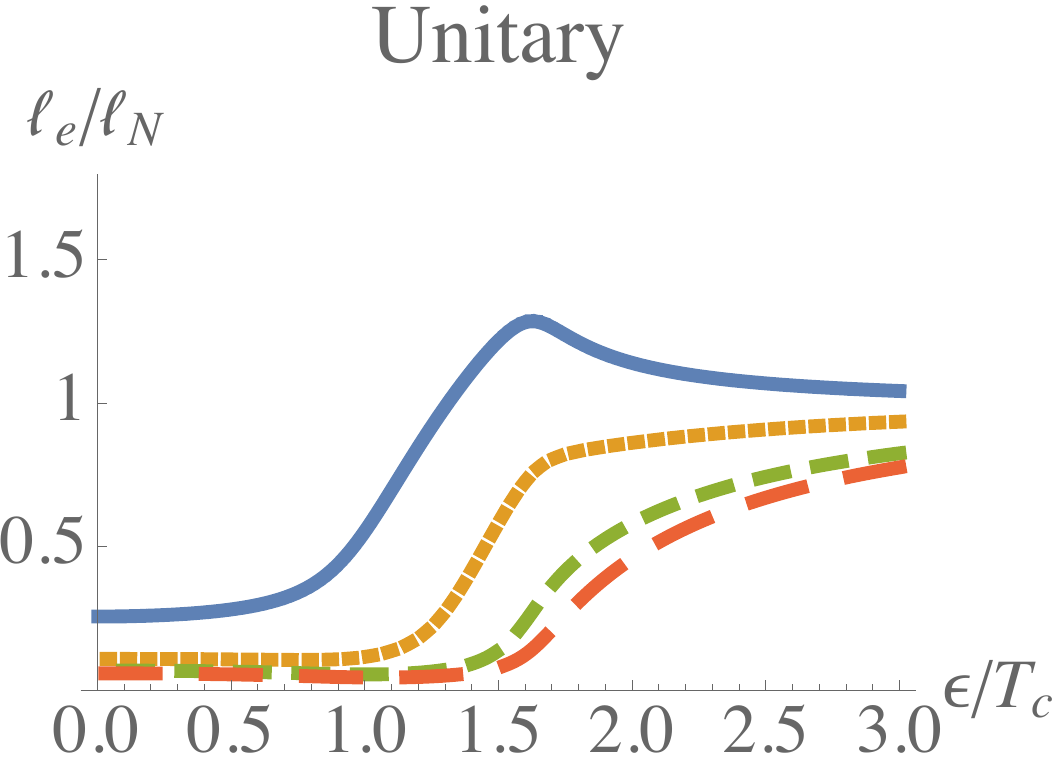}

  \includegraphics[width = 0.45\linewidth]{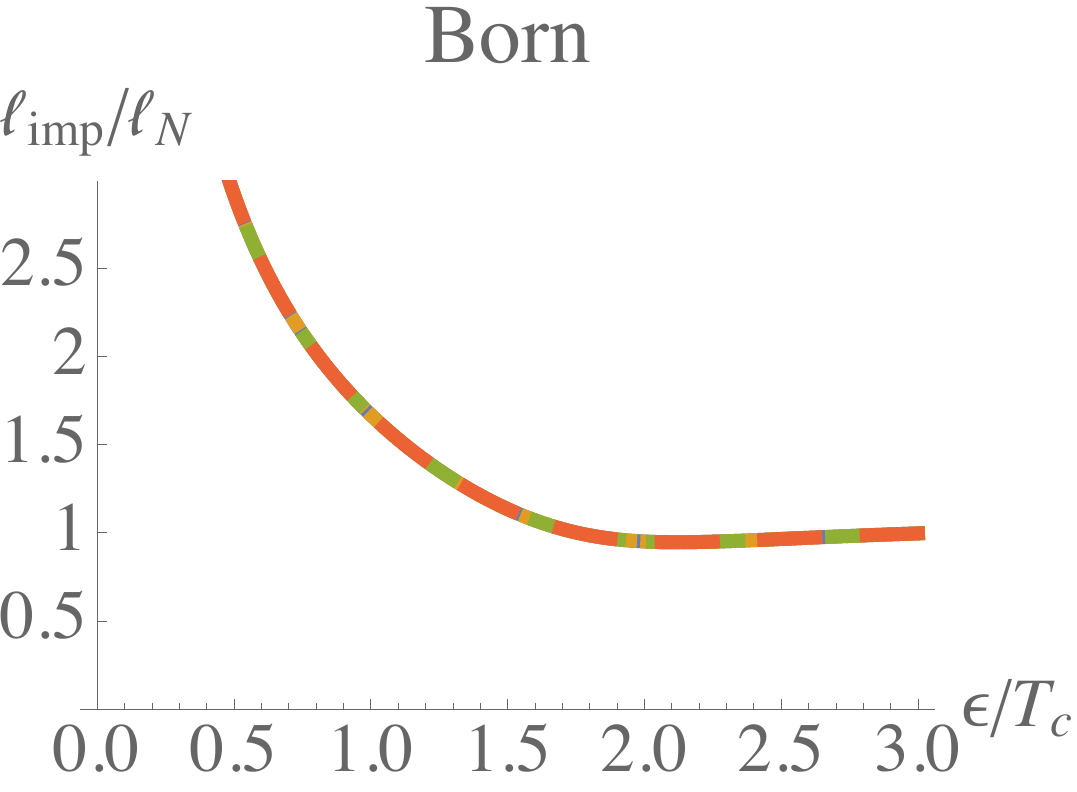} 
\includegraphics[width = 0.45\linewidth]{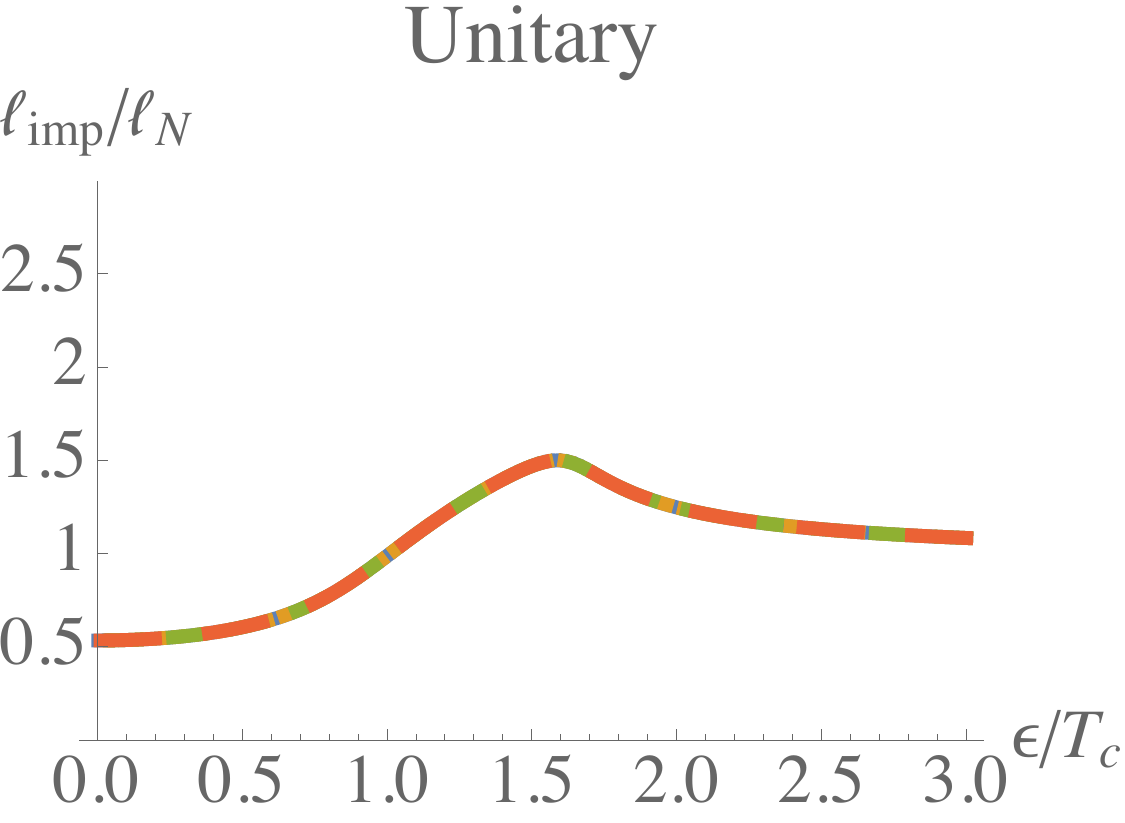}
\caption{\label{fig:uniform} 
\ %
Spectral and transport properties of 
a uniform d-wave superconductor. 
Angle resolved DoS, $N(\hp, \epsilon)/N_F$ (top row),   mean free path
$\ell_e(\hp, \epsilon)/\ell_N$ (middle row) and impurity scattering length  $\ell_{\rm imp}(\epsilon)/\ell_N$ 
(bottom row) are plotted in Born and Unitary limits, 
for the normal state mean free path $\ell_{ N} \approx 10\, \xi$, where  $\ell_{\rm imp}=v_F/2 \im[\Sigma_{\rm imp}] $. 
Different curves represent different momentum directions spanning the d-wave clover from a node to antinode (solid
blue to red dashed lines), as shown in inset. 
In unitary limit, the low energy impurity band in DoS is large, and   the mean free path is reduced by enhanced impurity
scattering. By contrast, in Born limit,  the impurity band  is  exponentially small and  the mean free path of nodal quasiparticles is longer.    
}
\end{figure}

The mean free path reflects the effectiveness 
of the scattering of quasiparticles by impurities. 
It depends on the concentration $\Gamma$ and strength $\delta$ of 
impurities, as well as on the available phase space for scattering, 
given by the properties of the order parameter $\Delta$. 
At low energy $\epsilon <W^B_\imp < \Delta$, in the Born limit, impurity
scattering is ineffective, $v_F/2\,\im[\Sigma^R_\imp]> \ell_{N}=v_F\tau_N 
=v_F /(2\Gamma \sin^2\delta )$, 
and it allows quasiparticle to travel long distance between scatterings producing 
large heat transport. 
By contrast, in the Unitary
limit, scattering is enhanced  $v_F/(2\,\im[\Sigma^R_\imp])< \ell_N $, i.e. low
energy QPs bind to impurities forming a wide impurity band. 

\begin{figure}[t!]
\includegraphics[width = 0.8\linewidth]{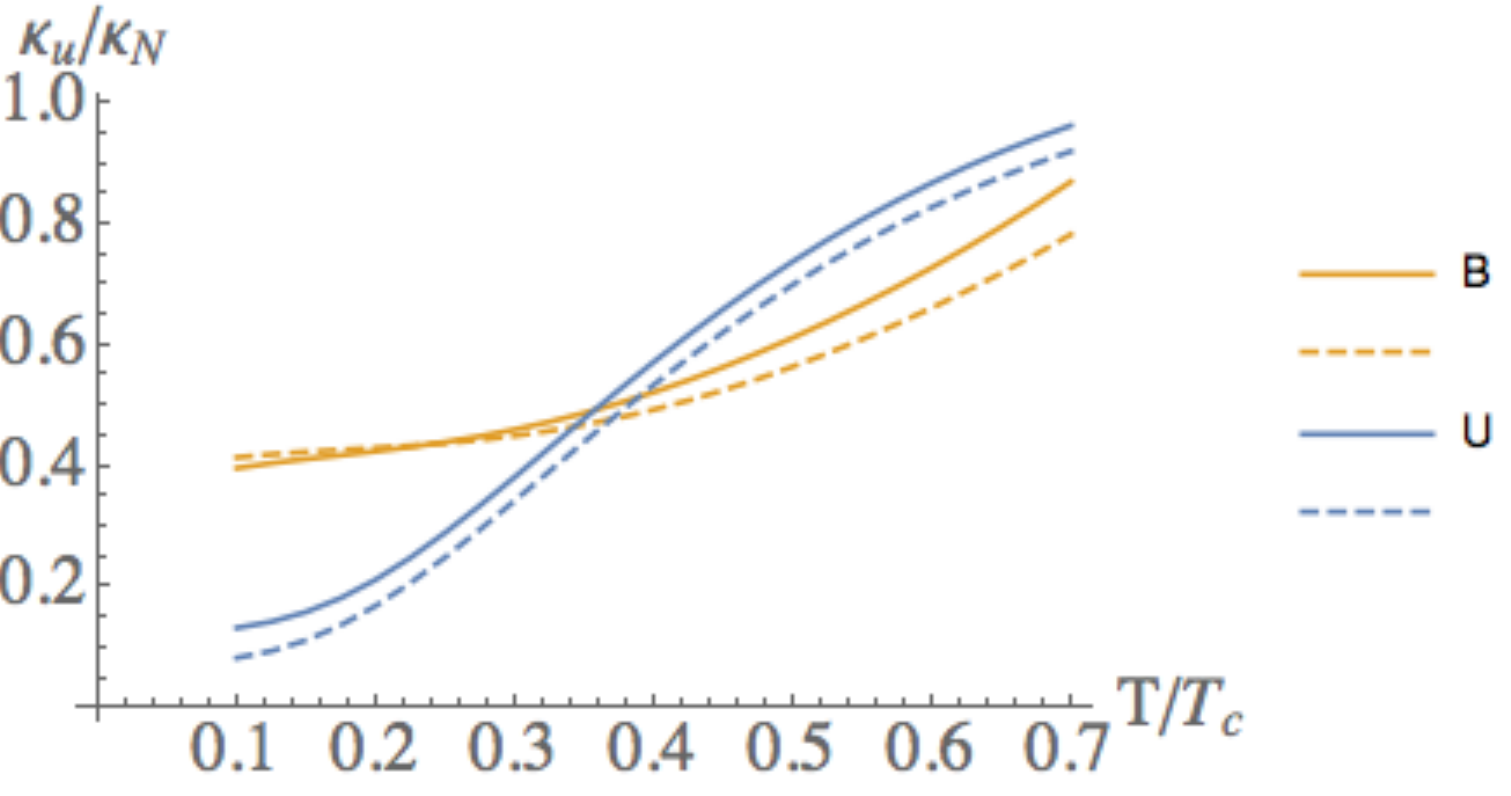}
\caption{ \label{fig:kUni} (Color online) 
Uniform thermal conductivity as a function of temperature. 
At low temperature $T \lesssim 0.3 T_c$ ($\epsilon \lesssim 0.6 T_c$) 
thermal conductivity 
in Born limit (green) is higher than that in Unitary limit (blue), 
indicating that it is dominated by large mean free path of quasiparticles. 
Solid and dashed lines correspond to mean free paths 
$\ell_N=\pi \xi/0.3 $ and $\ell_N=\pi \xi/0.2  $ respectively. 
} 
\end{figure}

Numerically, we find that thermal transport properties are mainly influenced by the behavior of 
scattering length $\ell_e(\hp,\epsilon)$ rather than that of density of states. 
In Fig.~\ref{fig:kUni},  we plot the temperature dependence of $\kappa_u(T)$, 
which we analyze using data from Fig.~\ref{fig:uniform}. 
At low-intermediate temperature $0.05 < T/T_c < 0.3$, 
corresponding to energies $W_{imp} \lesssim \epsilon < 0.6 T_c$ the DoS
in Born limit is small $ N^B(\epsilon ) < N^U(\epsilon)$, 
while $\ell_e^B \gg \ell_e^U$,
producing $\kappa_u^B(T)>\kappa^U_u(T)$.  
At higher energy and temperature  $0.4 < T/T_c$, $\epsilon > 0.8 T_c$ 
the result is reversed $\kappa_u^B(T)<\kappa_u^U(T)$, 
again in agreement with the increase of $\ell_e^U > \ell_e^B$ while 
having about the same values for the DoS in this energy interval. 
In the very low temperature limit, $T\ll W_\imp$, 
DoS and scattering effects exactly  cancel each other, 
producing the  universal limit for heat
conductivity, where it does not depend on the disorder properties.\cite{Lee1993universal,Graf1996,Taillefer1997universal}

\section{Heat conductivity of a clean constriction \label{app:clean}}

\begin{figure}[t!]
   (a) \hspace{0.4\linewidth} (b) \hspace{0.6\linewidth}   \\
\includegraphics[width =0.45\linewidth]{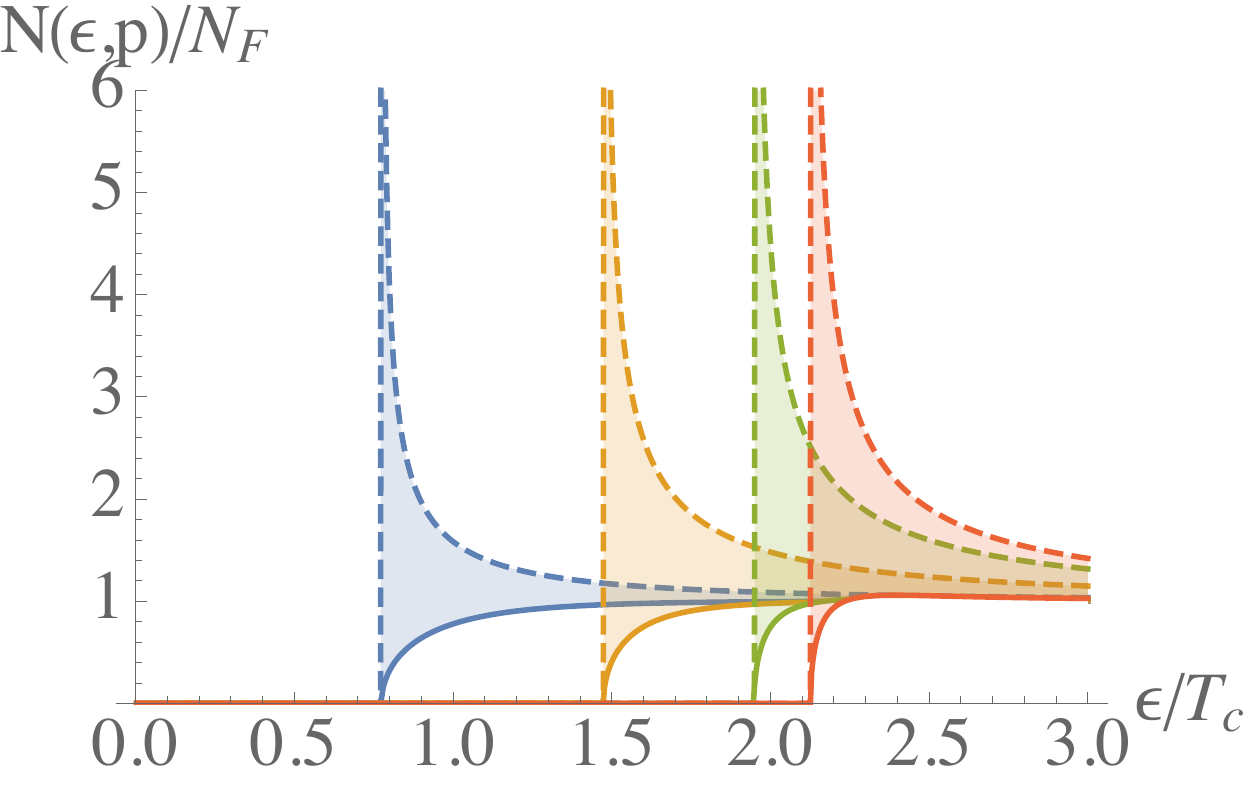}
\includegraphics[width =0.45\linewidth]{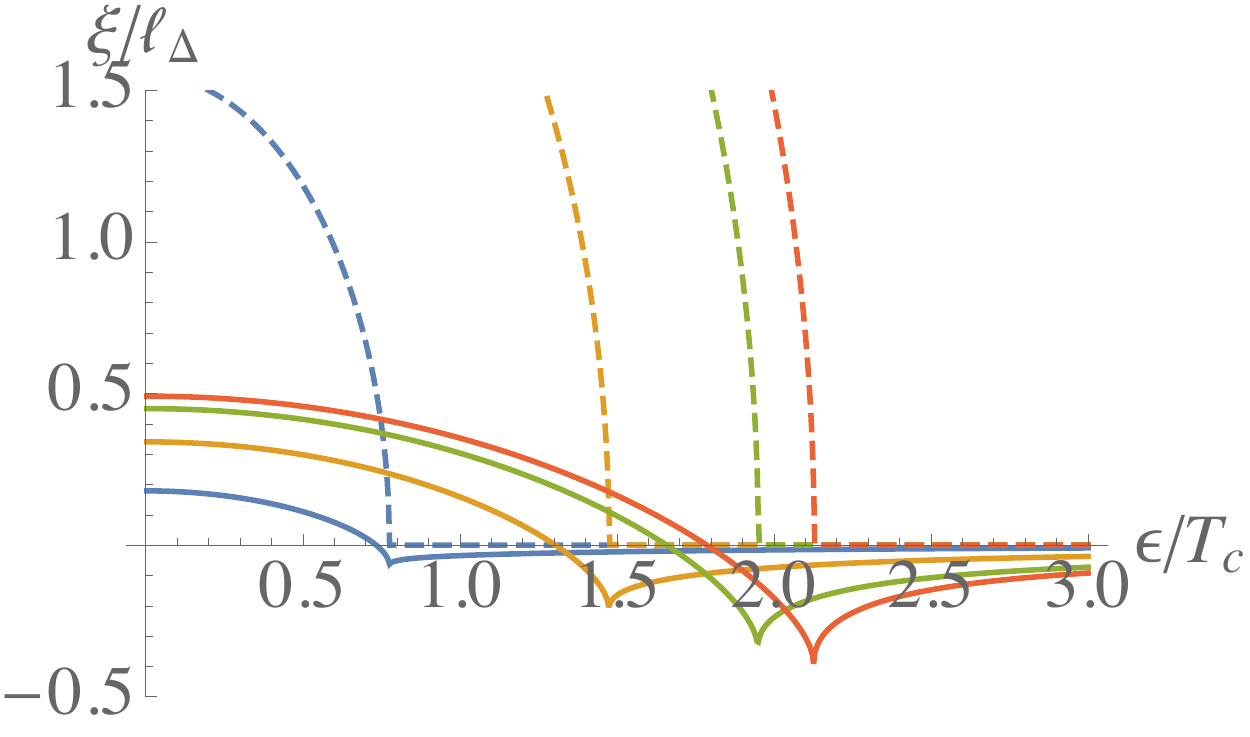} \\ 
     (c) \hspace{0.4\linewidth} (d) \hspace{0.6\linewidth}
     \includegraphics[width =0.5\linewidth]{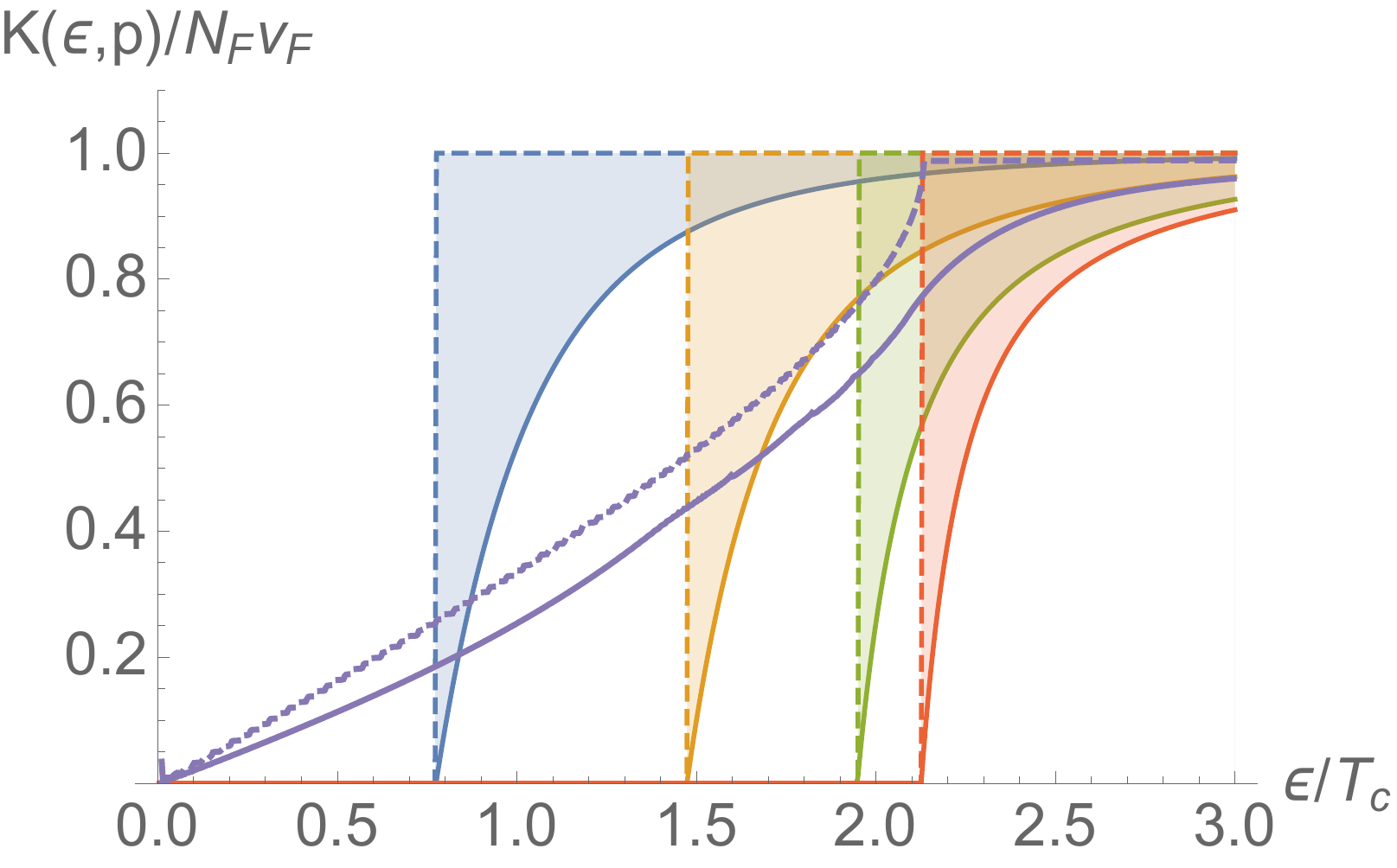}   
\includegraphics[width =0.38\linewidth]{clover.pdf}  \\
\caption{ \label{fig:clean} 
Thermal transport properties of a clean superconductor across a single domain wall (solid lines) compared agains uniform
superconductor (dashed lines). 
(a) Local DoS for momenta directions shown in (d) at the domain wall $N(\hp, \epsilon, x)$ with part of the spectral weight (shaded area) 
moved from continuum states into zero-energy bound states, that form a very sharp peak not resolved on this scale. 
(b)  The Andreev reflection length scale $\ell_\Delta(\epsilon, \hp, x)$. 
In uniform superconductor it is infinite for above-gap energies 
$ 1/\ell_\Delta \left(|\epsilon| > |\Delta(\hp)|\right) =0 $,
while at the domain wall it is finite for all energies and even changes sign. 
(c) Kernel of the heat current $K(\hp,\epsilon)$ for four momentum directions and integrated over the Fermi surface. 
With the domain wall the kernel $K(\epsilon, \hp)<1$ 
is suppressed due to Andreev reflection. 
}
\end{figure}

In this appendix, we evaluate    heat  transport properties  of a spin-singlet superconducting constriction without
impurities and discuss the role of Andreev reflection processes. 
The constriction can be thought of as a narrow bridge connecting two large reservoirs, that are assumed to be 
in equilibrium at temperature $T\pm dT$ ($dT\ll T,T_c$). We define  the conductance of the clean constriction  as $G= I_h/( 2dT )$. 
The global phases of the superconducting order parameter in the reservoirs $\Delta(\hp)\exp(i\varphi_{L,R})$ is set to
$\varphi_{L,R}=0,\pi$.  
The constriction is assumed to be long and narrow, so we neglect the edge effects. 
In linear response, the energy transport is governed by Eqs. \refe{eq:gamma} and \refe{eq:xAlin}, 
with $\sigma_\imp=0$. At boundaries,  $\gamma(\pm L, \hp_x\lessgtr0, \epsilon)$ is given by \refE{bc:gamma} and we take
    \beq
	x^a_{R/L} = x^a(\pm L, \hp_x\lessgtr 0, \epsilon)=  \partial_T \Phi_0 \,(\mp dT)
	(1+ \gamma_u^R\tilde \gamma_u^A) ,
    \eeq
which conveniently describes junctions between reservoirs that have negligible heat currents inside. 
This is different from the boundary condition \refe{bc:xa} that was aimed at describing a
continuous flow of heat. 
   
The order parameter $\Delta(x)$ and   $\gamma_0(x, \hp, \epsilon)$ are  self-consistently determined throughout the 
constriction. 
From equilibrium $\gamma_0(x, \hp, \epsilon)$, using \refE{eq:gamma}, one can find analytic solution for the distribution 
function along the constriction: 
  \begin{align}
  \begin{split}
   x^a(x, \hp_x > 0, \epsilon )= t(x, \hp, \epsilon) \; x^a_{L},  
   \\
   x^a(x, \hp_x < 0 , \epsilon )= t(x, \hp, \epsilon) \; x^a_{R},  
   \end{split}
   \label{eq:xaclean}
  \end{align}
where 
   \beq
   t(x, \hp, \epsilon)= \frac{1-|\gamma^R_0(x,   \hp , \epsilon)|^2}{1-|\gamma^R_u(\hp, \epsilon)|^2}, 
   \eeq 
plays the role of a transmission coefficient ($|t|<1$). 
In a uniform superconductor energy is perfectly transmitted  $ | t(\epsilon, \hp)|=1$. 
However, with a domain wall, one has $|t|\leq1$, i.e.  energy is not fully transmitted. This is interpreted  as a partial
Andreev reflection of incident quasiparticles from the spatially varying profile of the order parameter. 
Inserting \refE{eq:xaclean} into heat current expression \refE{eq:qLinResp}, we can express the conductance as 
       \beq 
	 G =  \int d\epsilon \;\epsilon \, \fsav{|\hp_x| \, K(\epsilon, \hp)}  \pder{\Phi_0}{T} \,,  
	 \eeq 
where the kernel $K(\epsilon, \hp)$  is 
       \beq
	 K(\epsilon, \hp)= N_F v_F\frac {(1- |\gamma_0^R(\epsilon, \hp, x)  |^2)(1- |\tilde\gamma_0^R(\epsilon, \hp, x) |^2)}
	 		{ |1+ \gamma^R_0(\epsilon, \hp, x)\tilde \gamma^R_0(\epsilon, \hp, x)|^2} \,. 
       \eeq
Again, because the energy flow is uniform, $ K(\epsilon, \hp)$ does not depend on position $x$, even though 
$\gamma_0^R$ does. 

In Fig. \ref{fig:clean} where we plot the heat current kernel together with 
the density of states and the Andreev reflection length $1/\ell_\Delta =2\,\im [\gamma^R \tilde\Delta]/v_F$ 
appearing in \refE{eq:xAlin}. 
For uniform order parameter (dashed lines) 
$K(\epsilon,\hp)=1$ for $\epsilon>|\Delta(\hp)|$, and is zero for energies below the gap
where there are no quasiparticle states. 
$\ell_\Delta(\epsilon, \hp)$ is finite for subgap states $\epsilon <\Delta(\hp)$, and infinite otherwise. 

At the center of domain wall $\ell_\Delta\left(\epsilon>\Delta(\hp),\hp, x\right)$ is finite (and can even be 
negative!) for the above-gap states, 
their spectral weight is moved into the ABS, and the amplitude of $K(\epsilon, \hp)$ is reduced, 
as shown by solid lines in Fig. \ref{fig:clean}. 
In the clean limit, the  conductance is reduced in  the presence of a single  domain wall, alike  the pinhole of perfect
transparency. \cite{ZhaoPRL2003}

\bibliographystyle{apsrev4-1} \bibliography{bibloheatcurrent}

%
%
%

\end{document}